\documentclass[aps,prd,twocolumn,nofootinbib]{revtex4-2}
\usepackage{graphicx}
\usepackage[utf8]{inputenc}
\usepackage{placeins}
\usepackage{amssymb, amsmath}
\usepackage{tabularx}
\usepackage[normalem]{ulem}
\usepackage{pbox}

\newcommand{\Ms}{M_{\odot}}

\newcommand{\rns}{\rho_{\rm sat}}

\newcommand{\Rsmall}{$R_{1.4}\approx11$~km\ }
\newcommand{\Rbig}{$R_{1.4}\approx14$~km\ }

\usepackage[dvipsnames, usenames]{xcolor}

\begin{document}

\title{Detectability of Finite-Temperature Effects From Neutron Star Mergers with Next-Generation Gravitational Wave Detectors}
\author{Carolyn A. Raithel,$^{1,2}$ Vasileios Paschalidis,$^{3,4}$}
\affiliation{$^1$School of Natural Sciences, Institute for Advanced Study, 1 Einstein Drive, Princeton, NJ 08540, USA}
\affiliation{$^2$Princeton Gravity Initiative, Jadwin Hall, Princeton University, Princeton, NJ 08540, USA}
\affiliation{$^3$Department of Astronomy and Steward Observatory, University of Arizona, 933 N. Cherry Avenue, Tucson, Arizona 85721, USA}
\affiliation{$^4$Department of Physics, University of Arizona, 1118 E. Fourth Street, Arizona 85721, USA}
\date{May 2023}

\begin{abstract}
Observations of the high-frequency gravitational waves (GWs) emitted by the hot and massive remnant of a binary neutron star merger will provide new probes of the dense-matter equation of state (EoS). We show that current uncertainties in the thermal physics can cause the emergent GW spectum to differ by a degree comparable to changing the cold EoS by $\pm\sim120$~m in the characteristic radius of a neutron star. Unless a very close binary neutron star merger takes place, these effects are unlikely to be measurable with current GW detectors. However, with proposed next-generation detectors such as Cosmic Explorer or Einstein Telescope, the effects can be distinguished
for events at distances of up to $\sim80$-200 Mpc, if the cold EoS is sufficiently well constrained. 
\end{abstract}

\maketitle

\section{Introduction}
Following a binary neutron star merger, if the masses of the initial neutron stars
are within a certain range, the remnant can survive
as a meta-stable, hypermassive or supramassive neutron star which will emit
high-frequency gravitational waves that are sensitive
to the details of the underlying equation of state (EoS)
 \cite{Baiotti:2016qnr,Paschalidis:2016vmz,Bernuzzi:2020tgt,Dietrich:2020eud,Radice:2020ddv}.
 Depending on the EoS, these post-merger gravitational waves (GWs)
 peak at frequencies of 2-4~kHz, with softer EoSs (which predict more
compact neutron stars) leading to higher peak frequencies, and
stiffer EoSs (which predict larger-radius neutron stars) generally leading to
lower peak frequencies \cite{Bauswein:2012ya,Bauswein:2011tp,Takami:2014zpa,Bernuzzi:2015rla}
\cite[see also][]{Breschi:2021xrx,Raithel:2022orm}.

Because the post-merger GWs peak at frequencies well above the $\sim100$~Hz regime
where the current generation of detectors are most sensitive,
measuring these signals with the existing network of GW detectors 
 will require either a very nearby source
($\lesssim$30~Mpc \cite{Clark:2015zxa}) and/or the development of
new data analysis techniques, e.g., for mode stacking of individual events \cite{Yang:2017xlf}. 
On the other hand, with
the construction of next-generation (XG) GW detectors, 
such as the proposed
Cosmic Explorer (CE) \cite{Reitze:2019iox}, Einstein Telescope (ET) \cite{Punturo:2010zz}, or NEMO 
\cite{Ackley:2020atn} detectors, these signals are expected to be detectable within $\lesssim1$~year of
operations \cite{Yang:2017xlf,Torres-Rivas:2018svp,Evans:2023euw}.

One key difference between the EoS constraints that have been placed using inspiral GWs
\cite[e.g.,][]{LIGOScientific:2018cki,Baiotti:2019sew,GuerraChaves:2019foa,Raithel:2019uzi,Chatziioannou:2020pqz,Annala:2021gom}
and what will be probed with the post-merger GWs in the near future,
is that whereas the inspiraling neutron stars are thermodynamically cold,
the post-merger remnant is significantly shock-heated, with thermal pressures that
can be a significant fraction ($\gtrsim$10\% at supranuclear densities) of the cold pressure (e.g., \cite{Oechslin:2006uk,Bauswein:2010dn,Sekiguchi:2011zd,Foucart:2015gaa,Palenzuela:2015dqa,Kastaun:2016yaf,Radice:2017zta,Raithel:2021hye,Raithel:2023zml}). As
a result, the emergent GW spectrum is sensitive not
only to uncertainties in our knowledge of the cold physics,
but also to our uncertain knowledge of the 
finite-temperature EoS 
\cite{Bauswein:2010dn,Figura:2020fkj,Raithel:2023zml,
Fields:2023bhs,Villa-Ortega:2023cps,Miravet-Tenes:2024vba}.

Finite-temperature effects have been shown to shift the spectrum
of post-merger GWs by up 
 to $\sim60-200$Hz for realistic thermal prescriptions 
 \cite{Raithel:2023zml,Fields:2023bhs,Villa-Ortega:2023cps,Miravet-Tenes:2024vba},
 with larger shifts of the peak frequency found when the cold
 EoS is relatively soft (i.e., predicts smaller radius neutron stars, leading to more
 violent collisions and stronger shock heating) \cite{Raithel:2023zml}.

In this work, we expand on our recent study of finite-temperature
effects in neutron star merger simulations \cite{Raithel:2023zml} and
we quantify the distinguishability of different prescriptions of the
thermal physics in the post-merger GWs. We compare the impact of these
thermal effects to a set of new simulations that explore the impact of
making small (few percent) variations to the cold EoS.  We find that,
for a baseline EoS model that predicts a characteristic neutron star
radius of \Rsmall, the uncertainty in the thermal physics leads to a
comparable change in the post-merger GWs as changing the cold EoS by
$\pm$120~m in $R_{1.4}$ or by $\pm$15 in the corresponding tidal deformabilty.
This level of uncertainty in the cold EoS is expected to
 be pinned down from inspiral GWs within
 one year of observations for an XG detector such as CE~\cite{Carson:2019rjx,Chatziioannou:2021tdi,Finstad:2022oni}. Thus, by combining
EoS constraints from the (cold) inspiral with a detection of the post-merger
GWs from the hot neutron star remnant, these results suggest that it will become possible
to constrain the \textit{finite-temperature} part of the EoS in the XG era,
at least for the thermal models considered here and assuming other 
second-order effects (e.g., out-of-equilibrium
effects \cite{Hammond:2022uua,Most:2022yhe,Espino:2023dei}) are sufficiently well
understood.

We quantify the distinguishability of thermal effects in the
post-merger GWs with current and upcoming GW detectors.
We find that thermal effects are easier to distinguish for a stiffer EoS
than for a softer EoS, even though there is significantly less heating 
for the stiff EoS and the shifts to the dominant peak frequencies are accordingly 
smaller. 

We conservatively estimate that the thermal effects studied here
could be resolved at distances of up to $\sim200$~Mpc  for a stiff (\Rbig) 
EoS and up to $\sim$80~Mpc for a soft (\Rsmall) EoS,
with the sensitivity of a 20~km Cosmic Explorer detector
that has been tuned to optimize post-merger sensitivity \cite{Srivastava:2022slt},
if the cold EoS is well known  (see also~\cite{Villa-Ortega:2023cps}).
With the design sensitivity of aLIGO, a merger would have to be within
$\lesssim10$~Mpc to measure these thermal effects. We compare these
prospects with the proposed sensitivity of ET and 
with a 40~km configuration of CE and determine that the 20~km 
``post-merger tuned" configuration of CE
leads to the best prospects for measuring finite-temperature effects
directly from the post-merger GWs. We note that a full
detectability analysis -- including a framework to  
measure the statistical significance of the
inferred thermal parameters -- would 
require a parameter estimation study, which is beyond the scope
of the present work.
Nevertheless, our results indicate that the
prospects are optimistic
for distinguishing these thermal prescriptions from 
neutron star mergers at realistic distances,
with the projections for all three of the 
proposed next-generation detectors.

\section{Numerical Methods}
In this work, we explore the observability of finite-temperature effects in the post-merger GWs
from a binary neutron star merger. To that end, we analyze the GW signals from a set
of simulations that were designed to systematically disentangle the relative roles of the uncertainties in the zero-temperature
and finite-temperature parts of the EoS. Some of these simulations were
previously presented in \cite{Raithel:2023zml}. 
We briefly review the key details of the set-up
here, but for further details see \cite{Raithel:2023zml} (hereafter, Paper I).

\subsection{Construction of finite-temperature equations of state}
In both the simulations of Paper I and the new simulations performed in this work,
we use the phenomenological framework of \cite{Raithel:2019gws} for calculating the EoS.
In this framework, the total pressure is decomposed into
a cold and thermal component, according to
\begin{equation}
P_{\rm total}(n, T, Y_e) = P_{\rm cold}(n, T=0, Y_e) + P_{\rm th}(n, T, Y_e)
\end{equation}
where $n$ is the baryon number density, $T$ is the temperature, and $Y_e$ is the electron fraction, 
which we fix to the initial $\beta$-equilibrium configuration. For the cold EoSs, we adopt
 two different generalized-piecewise polytropic EoSs \cite{OBoyle:2020qvf,Raithel:2023zml}, which predict 
 characteristic neutron star radii of \Rsmall and 14~km.
 
We additionally construct modified versions of the softer (\Rsmall) model, which
vary from the baseline model by $\Delta R_{1.4}=-120,-54,$ and $+116$~m 
(or equivalently, by up to $\pm5\%$ in the cold pressure at supranuclear densities;
see Appendix~\ref{sec:appEOS} for details).
This range of cold EoSs is motivated by projections that
Cosmic Explorer will measure the 
neutron star radius to within 50-200~m   
within one year of inspiral observations \cite{Chatziioannou:2021tdi,Finstad:2022oni}. 
Thus, this set of cold EoSs brackets the degree of variation that is expected to
be constrained in the XG era.

\begin{figure*}[!ht]
\centering
\includegraphics[width=0.9\textwidth]{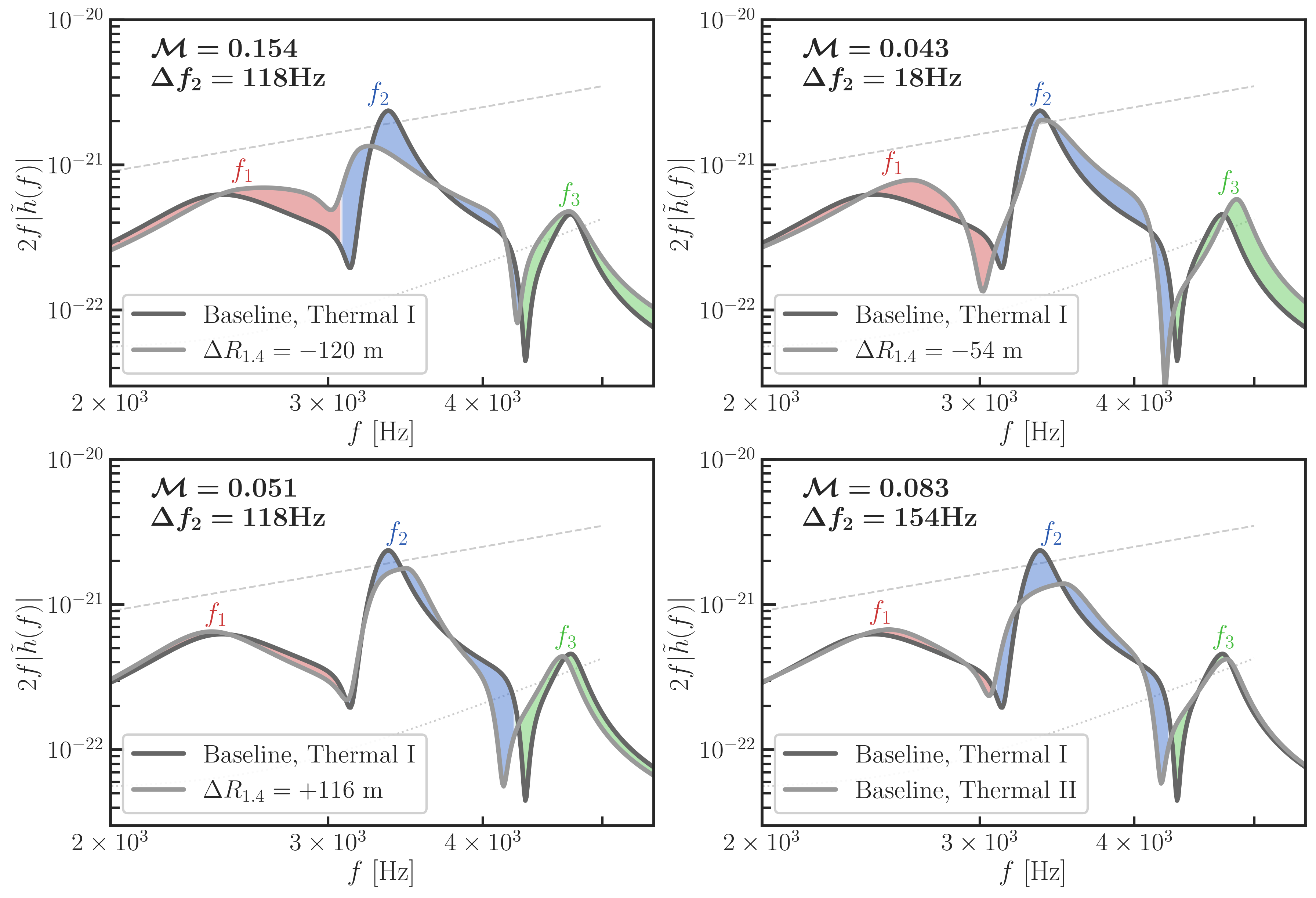}
\caption{\label{fig:spectra} 
Post-merger GW spectra for a subset of the
\Rsmall evolutions. We show here the best-fit templates for each simulation, based on a model comprising three skew-Lorentzian peaks (see Appendix~\ref{sec:appGW} for details). The dashed and dotted lines indicate the sensitivity of aLIGO at design sensitivity \cite{aLIGOsensitivity} and CE in the 20~km configuration tuned for post-merger sensitivity \cite{Srivastava:2022slt}, respectively. Spectra include the dominant $\ell=m=2$ mode and assume a face-on orientation for a source directly overhead and at a distance of 40 Mpc. The top row and bottom left panels show the impact of varying the cold EoS by -120, -54, and +116~m in radius (or, equivalently, scaling the supra-nuclear pressure by up to $\pm$5\%). The lower right panel shows the impact of varying the $\alpha$-parameter of the $M^*$ thermal prescription. The color-coding is included to help visually distinguish the three spectral peaks (labeled $f_1, f_2$ and $f_3$) and to highlight the mismatch between each pair of spectra. }
\end{figure*} 

To each cold EoS, we attach one of four prescriptions for the thermal pressure, calculated using the ``$M^*$-framework" of \cite{Raithel:2019gws}, which captures the leading-order effects of degeneracy in the thermal prescription by parametrizing the particle effective mass, $M^*$. The framework has been shown 
to  accurately recreate full EoS tables both analytically \cite{Raithel:2019gws} and in the context of merger simulations \cite{Raithel:2022nab}. The four thermal prescriptions used here correspond to four choices of $M^*$-parameters, including density parameters $n_0 \in (0.08, 0.22)$~fm$^{-3}$, which describe the density at which the effective mass starts to decay from its vacuum rest-mass, and power-law parameters $\alpha \in (0.6,1.3)$, which characterize the rate at which $M^*$ decreases with density. This range of parameters was found previously to bracket the range of best-fit values for realistic calculations of the effective mass function for commonly-used finite-temperature EoSs \cite{Raithel:2019gws}. We refer to these parameter sets as Thermal Cases I-IV in the following.

\begin{table}
\centering
\begin{tabular}{ccccc}
\hline \hline
Cold EoS     & \pbox{20cm}{\vspace{0.05cm}Thermal \\Case\vspace{0.05cm}}  &  $M^*$-parameters  &  $\langle P_{\rm th}/P_{\rm cold} \rangle$  &   $f_2$ [kHz] \\
 \hline
       & I               & (0.08~fm$^{-3}$, 0.6) &  0.33 &  3.35  \\
\Rsmall ~~ & II  & (0.08~fm$^{-3}$, 1.3) &  0.27   &  3.50  \\
       & III            & (0.22~fm$^{-3}$, 0.6) &  0.34   &  3.33  \\
       & IV   	        &  (0.22~fm$^{-3}$,  1.3) & 0.40  &  3.48  \\
\hline
       & I             & (0.08~fm$^{-3}$, 0.6)&  0.12   &  2.62  \\
\Rbig ~~ & II   &  (0.08~fm$^{-3}$, 1.3)  &  0.12   &  2.57  \\
       & III            & (0.22~fm$^{-3}$, 0.6) &  0.10   &  2.63  \\
       & IV  	        &  (0.22~fm$^{-3}$,  1.3) &  0.14   &  2.58  \\
\hline \hline
\end{tabular}
\caption{Summary of the simulations varying the thermal treatment, for two different cold EoSs. From left to right, the columns indicate the cold EoS, the thermal prescription, the $M^*$ parameters ($n_0$ and $\alpha$), the density-weighted average the thermal pressure relative to the cold pressure, and the peak GW frequency. The averages are computed including all matter with densities $\rho \ge \rho_{\rm sat}$.}
  \label{table:thermal}
\end{table}

\begin{table}
\centering
\begin{tabular}{lccccc}
\hline \hline
Cold EoS   & $\Lambda_{1.4}$  && $\langle P_{\rm th}/P_{\rm cold} \rangle$  &  $f_2$ [kHz] \\
 \hline
 $\Delta R_{1.4}=-120$~m & $\Delta \Lambda_{1.4}=-16$ &  &  0.36   &    3.18  \\
 $\Delta R_{1.4}=-54$~m & $\Delta \Lambda_{1.4}=-7$ & &  0.28   &  3.35  \\
  Baseline [ $R_{1.4}=11.1$~km ] & $\Lambda_{1.4}=230$  & &   0.33  &   3.35    \\
 $\Delta R_{1.4}=+116$~m &  $\Delta \Lambda_{1.4}=+14$   &&   0.28 &   3.46 \\
\hline \hline
\end{tabular}
\caption{Summary of simulations varying the cold (\Rsmall) EoS. The first two columns indicate the deviations, relative to the baseline model, for the radius and tidal deformability of a 1.4~$\Ms$ neutron star.
All simulations use thermal prescription I.
The baseline case is identical to the \Rsmall entry from Table~\ref{table:thermal}, with
$M^*$-parameters (0.08 fm$^{-3}$, 0.6).}
  \label{table:cold}
\end{table}

We attach each of the four thermal prescriptions to the \Rsmall and \Rbig cold EoS models (for a total of 8 finite-temperature EoSs), to explore the sensitivity of the merger to the choice of thermal treatment. To the set of modified \Rsmall EoSs, we attach a single thermal treatment (Case I), to explore the sensitivity of the merger to small variations in the cold EoS with a fixed thermal treatment. We summarize the EoSs thus constructed in Tables~\ref{table:thermal} and \ref{table:cold}. 

\subsection{Initial conditions and numerical setup}
\label{sec:IC}
For the new simulations performed in this work, the numerical set-up is identical to that used for the \Rsmall evolutions in Paper I \cite{Raithel:2023zml}. In summary,
for each EoS, we perform a simulation of an equal-mass (1.3+1.3$\Ms$) binary neutron star merger in full general relativity using the code of \cite{Duez:2005sg,Paschalidis:2010dh,Etienne:2011re,Etienne:2015cea}, as it has been recently extended in \cite{Raithel:2021hye,Raithel:2022san}. The initial neutron stars are constructed using \texttt{Lorene2} \cite{Lorene}, and are cold, irrotational, un-magnetized, and placed at an initial coordinate separation of 40~km.  

 Our simulations use nine spatial refinement levels, which are separated by a 2:1 refinement ratio. The grid spacing on the innermost refinement level corresponds to $\Delta x=140$~m, such that the coordinate diameter of the initial neutron stars are covered by $\sim$125 grid points (where the initial coordinate radius of this model is 8.76~km along the x-direction). For the \Rbig EoSs, the grid spacing on the innermost refinement level is $\Delta x=195$~m, for the same effective grid coverage across the initial neutron stars (with initial coordinate radii of 12.10~km) .

\section{Results}
For all cases, we evolve the neutron stars through the final 3-4.5 orbits of the inspiral, through the merger, and for the first 12~ms post-merger 
(for the \Rsmall models) and the first 20~ms post-merger 
(for the \Rbig models), in order to extract the properties of the post-merger remnant and the emergent GW signal. We note that the unequal lengths
of the simulations are necessary to capture the salient features
of the post-merger GW emission, which are more temporally-extended
for the stiffer cold EoS. Spectrograms of the GW signals for
these models, and additional
discussion of their temporal differences, can be found
in Appendix~\ref{sec:appSpec}.

In this work, we focus primarily on the spectra of post-merger GWs, as the key observable feature from such events. Nevertheless, to give a sense of the different degrees of heating experienced in each evolution, Tables~\ref{table:thermal} and \ref{table:cold} also report the average thermal pressure relative to the cold pressure, $\langle P_{\rm th}/P_{\rm cold} \rangle$, within the late-time remnants, for each EoS. These averages are density-weighted and include all matter with densities $\ge \rns$ (where $\rns=2.7\times10^{14}$~g/cm$^3$ is the nuclear saturation density), in order to highlight the heating in the dense-matter core.
The summary features from the sequence of EoSs that vary in their thermal treatment, while holding the cold EoS fixed, are shown in Table~\ref{table:thermal} (for further analysis of these evolutions, see Paper I). The summary features from the new simulations, for the systematic variations to the cold \Rsmall EoS, are summarized in Table~\ref{table:cold}.

\subsection{Post-merger GW spectra}
We extract the GW signal from each of our simulations using the Newman-Penrose formalism. The resulting spectra of GWs are characterized by three main spectral peaks, as is typical of such simulations \cite[e.g.,][]{Stergioulas:2011gd,Takami:2014zpa,Rezzolla:2016nxn,Baiotti:2016qnr, Paschalidis:2016vmz,Bauswein:2019ybt}. However, the spectra also exhibit small-scale noise on top of these spectral peaks, which can originate either from numerical error or from turbulent motions in the remnant following the merger. This high-frequency noise can artificially inflate estimates of the differentiability between the spectra. In order to minimize the impact of this noise on our analysis, we fit the (frequency-domain) spectra with smooth templates that are constructed to capture the dominant spectral features of the simulations. Our templates thus comprise three Lorentzian profiles, which allow for non-zero skew in each peak. 
We perform a least-squares minimization to fit the parameters of
these template models to the spectra (for details, see Appendix~\ref{sec:appGW}).
We find that the best-fit templates are able to reliably reproduce the raw spectra, while reducing the dependence of our subsequent analysis on the small-scale noise. In particular, for estimating the distinguishability of the spectra, we find that using the templates always provides a more {\it conservative} estimate than would be estimated from the raw spectra. For additional discussion, see Appendix~\ref{sec:appGW}.

\begin{figure}[!ht]
\centering
\includegraphics[width=0.45\textwidth]{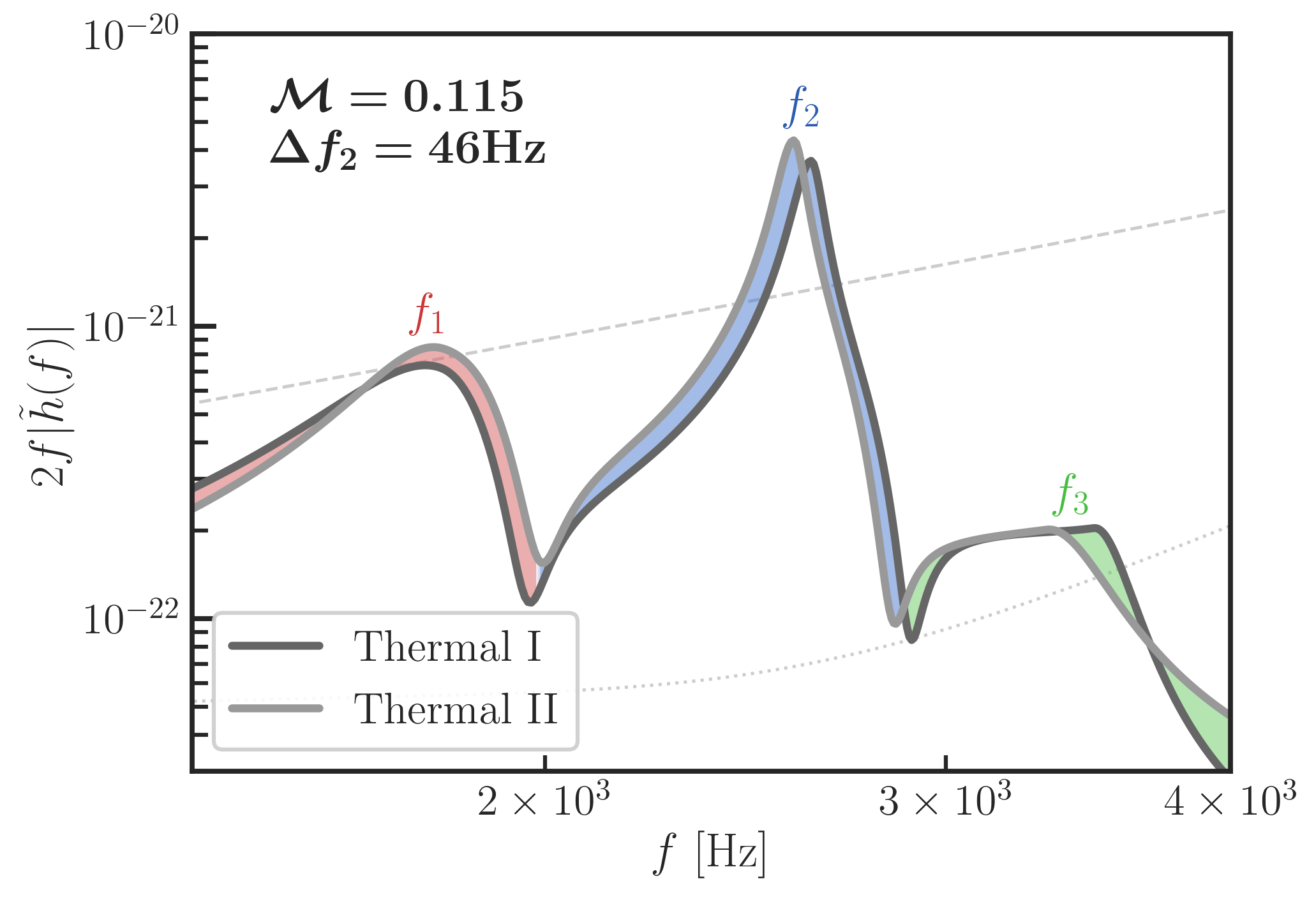}
\caption{\label{fig:spectra_R14} 
Same as the bottom right panel of Fig.~\ref{fig:spectra}, but for the scenario in which the thermal treatment is varied (from Case I to Case II) and the cold EoS is held fixed to the \Rbig model. }
\end{figure}

We show the resulting, best-fit templates to the post-merger GWs for several pairs of EoSs in Figs.~\ref{fig:spectra} and \ref{fig:spectra_R14}.
In the top row and bottom left panels of Fig.~\ref{fig:spectra}, we show the spectra from the simulations performed for the new
cold EoSs, and include the baseline \Rsmall model with the same thermal treatment (Case I) for
reference. In the bottom right panel of Fig.~\ref{fig:spectra}, we compare two choices for
 the thermal treatments (Case I and II), for which the cold EoS is held fixed to the \Rsmall model. Figure~\ref{fig:spectra_R14} shows the same comparison of the Case I and II thermal treatments, but for the evolutions in which the cold EoS is fixed to the \Rbig model.

These figures label the three dominant peaks ($f_1$, $f_2$, and $f_3$)
 and color-codes these spectral peaks in red, blue, and green, for visual clarity. In addition, 
 in the top corner of each subplot, Figs.~\ref{fig:spectra} and \ref{fig:spectra_R14}
 also report the differences in peak frequency, $f_2$, between each
 pair of EoSs shown, as well as the relative mismatch of the two spectra.
 We calculate the mismatch as
 \begin{equation}
 \label{eq:M}
\mathcal{M} = 1 - \frac{<h_1 | h_2>}{\sqrt{<h_1|h_1><h_2|h_2>}},
\end{equation}
where $<h_i|h_j>$ indicates the noise-weighted
inner product between the two waveforms, computed in the frequency
domain according to
\begin{equation}
\label{eq:innerP}
<h_i| h_j> = 4 \Re \int_{f_{\rm min}}^{f_{\rm max}} \frac{ \tilde{h}_i(f) \tilde{h}^*_j(f) }{S_n(f)} df.
\end{equation}
In this equation, $S_n(f)$ is the power spectral density of the detector noise, which we take (unless otherwise noted) to be that of CE in the 20~km configuration tuned for post-merger sensitivity \cite{Srivastava:2022slt}.
The asterisk in eq.~\ref{eq:innerP} indicates the complex conjugate of the strain and
 the integrals are computed over a frequency range $f_{\rm min}=2$~kHz (1.4 kHz) 
 to $f_{\rm max}=5.5$~kHz (4 kHz), for the \Rsmall (\Rbig) EoSs.
 This choice of frequencies approximately brackets the post-merger signal,
which we define as beginning just below the instantaneous GW frequency at merger
 and ending where the spectral amplitude decreases by $\sim100\times$ from the peak.
 We report  the mismatches after minimization
 with respect to phase and time shifts \cite{Lindblom:2008cm,McWilliams:2010eq} and using the best-fit spectral templates, which we find leads to mismatches that are typically smaller than ($\sim0.2-0.8\times$) the mismatches calculated from the raw spectra. As such, the templates provide a conservative, lower-limit estimate of  distinguishability.
 
 For reference, we include in Appendix~\ref{sec:appGW} the best-fit templates
 and corresponding mismatches for two simulations performed at two different resolutions 
 (the higher of which matches the effective resolutions of the \Rsmall and \Rbig evolutions),
 for an EoS with an intermediate stiffness and choice of $M^*$-parameters.
 For the two resolutions studied, we find only a small difference in the peak frequencies 
 of 26~Hz and a mismatch  between the best-fit templates
 of 0.037, which provides an approximate reference baseline for the following discussion. 
  
In comparing pairs of models that differ by small degrees in the cold
EoS, we find differences of up to 118~Hz in the peak frequency,
with the largest difference for the cases of either $\Delta R_{1.4}=-120$~m or +116~m, compared
to the baseline model. Of  the models considered in 
Fig.~\ref{fig:spectra},
the case of $\Delta R_{1.4}=-120$~m also has the largest mismatch with respect to the baseline model,
 with significant misalignments visible around not only the dominant
(second) peak, but around the first peak as well. 
For the case of $\Delta R_{1.4}=-54$~m, we find negligible differences
 in the peak frequency, but that there is still a significant mismatch, due
 to the differences around the first peak.
 At the other extreme, for the $\Delta R_{1.4}=+116$~m case, we find that
 the first peaks are very closely aligned, and that the second peak
 dominates the mismatch.
 
  The magnitude of these spectral shifts from small changes to the cold EoS
 are comparable to -- and in some cases smaller than --
 the impact of changing the thermal treatment.
 The lower right panel of Fig.~\ref{fig:spectra} shows
 one representative comparison of thermal effects, for the \Rsmall cold EoS with
 two different choices of the $M^*$-parameter $\alpha$, which was found in Paper I to govern the
 location of the peak frequency of the post-merger spectra.
 The difference in $f_2$ for these two thermal treatments is 154~Hz, slightly
 larger than changing the cold EoS by $\Delta R_{1.4}=\pm120$~m. 
 We note
 that this effect is typical of the changes to $f_2$ found for
 varying values of the thermal parameter, $\alpha$. We show additional examples
 between the other thermal cases in Appendix~\ref{sec:appGW}.
 Moreover, we find that the different thermal prescriptions (in the bottom right of Fig.~\ref{fig:spectra}, as well for those shown in Appendix~\ref{sec:appGW})
 lead to a comparable (and in some cases) larger
 mismatch than changing the cold EoS by -54~m to +116~m. 
Notably, the misalignment is visibly dominated by the dominant 
(second) peak when the thermal prescription is changed. 

 For the case of a stiffer cold EoS (\Rbig), the differences in peak frequency are 
 generally smaller ($\lesssim60$~Hz), though there are still significant mismatches 
($\mathcal{M}\sim0.1-0.26$), depending on the choice of the thermal parameter, $\alpha$ 
(see Fig.~\ref{fig:spectra_R14} and Appendix~\ref{sec:appGW}). This further highlights
that small differences in $f_2$ can correspond to spectra with large mismatches, and that
the thermal effects can have a {\it signifcant} effect on the post-merger spectra.

\subsection{Frequency dependence of the spectral mismatches}
 
\begin{figure}[!ht]
\centering
\includegraphics[width=0.45\textwidth]{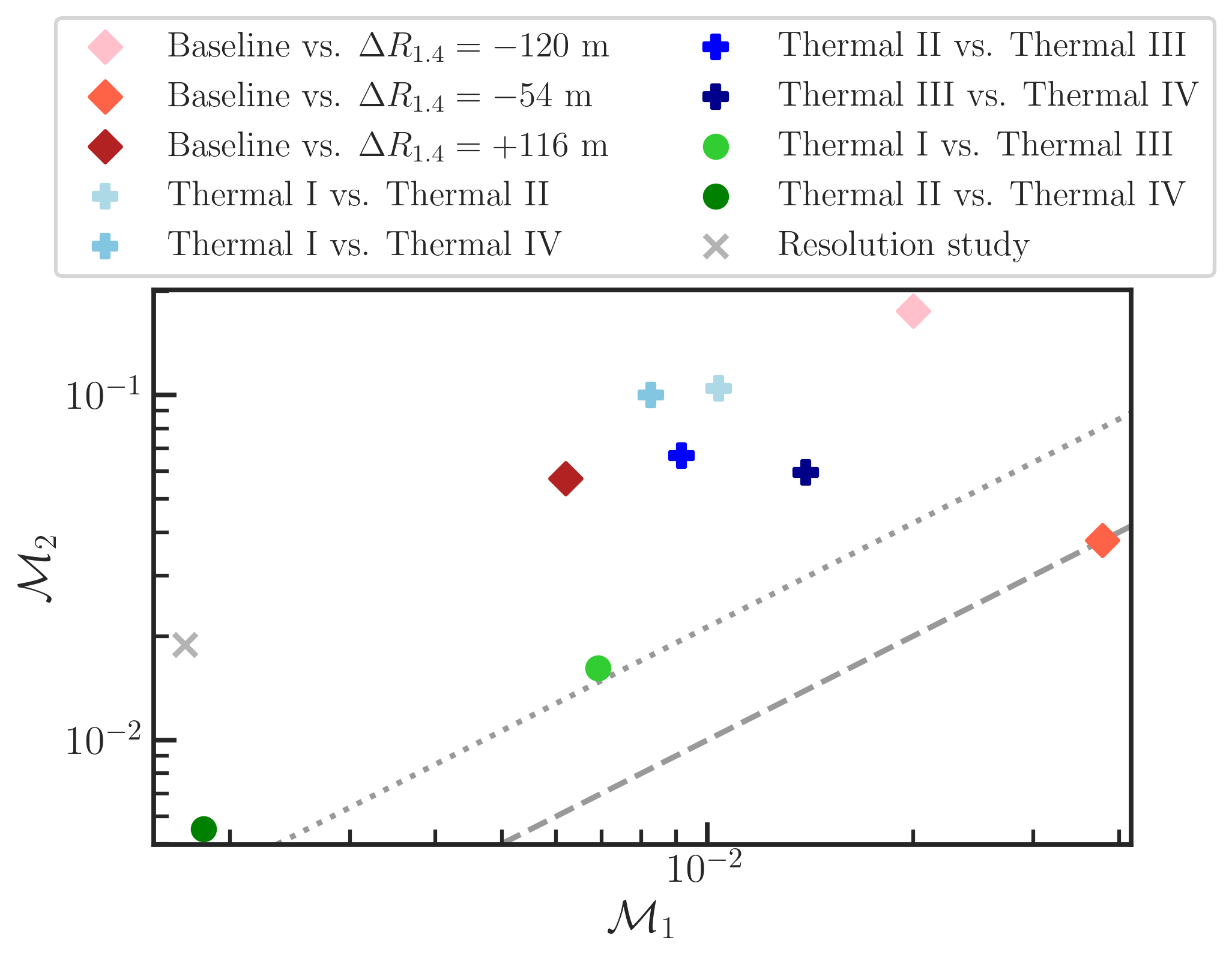}
\includegraphics[width=0.45\textwidth]{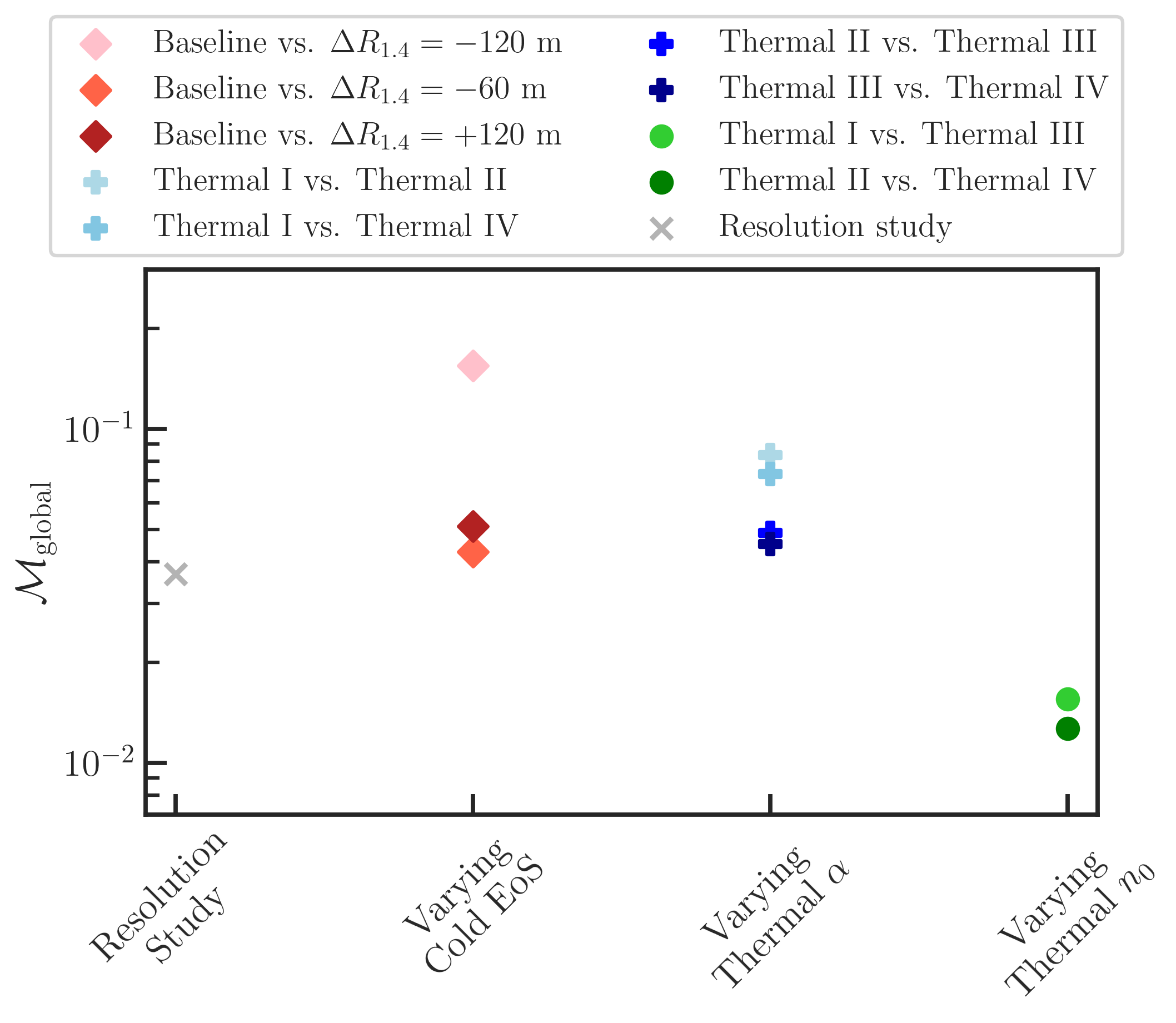}
\caption{\label{fig:peaks} Top: Mismatch of the first two peaks (corresponding to the red and blue shaded regions from Fig.~\ref{fig:spectra}) 
for various pairs of \Rsmall EoSs. Bottom: Global mismatches (calculated from 2-5.5~kHz) for the same models.
Comparisons for which the cold EoS has been varied are shown in red diamonds. Comparisons for which the cold EoS is held fixed (to \Rsmall) and the thermal treatment have been varied are shown in blue crosses (for different choices of the $M^*$-parameter $\alpha$) and green circles (for different choices of the $M^*$-parameter $n_0$). The gray x indicates the mismatches for two different resolutions and an EoS of intermediate stiffness ($R_{1.4}\approx12$~km) and choice of $M^*$-parameters. Finally, the dashed gray line in the top figure indicates the line of equal mismatches while the dotted gray line indicates where the horizon distances (eq.~\ref{eq:dhor}) would be equal.}
\end{figure}

\begin{figure}[!ht]
\centering
\includegraphics[width=0.45\textwidth]{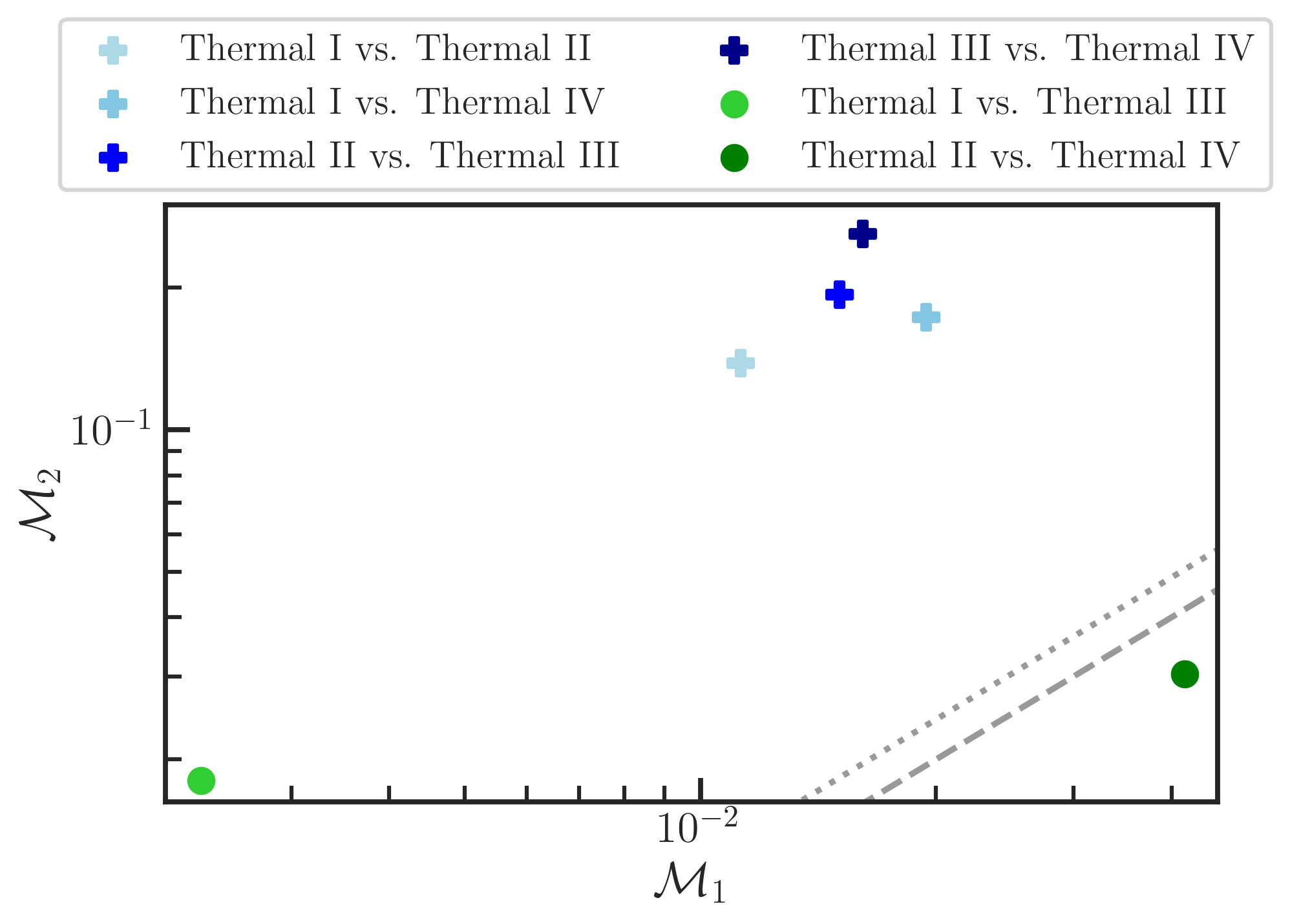}
\includegraphics[width=0.45\textwidth]{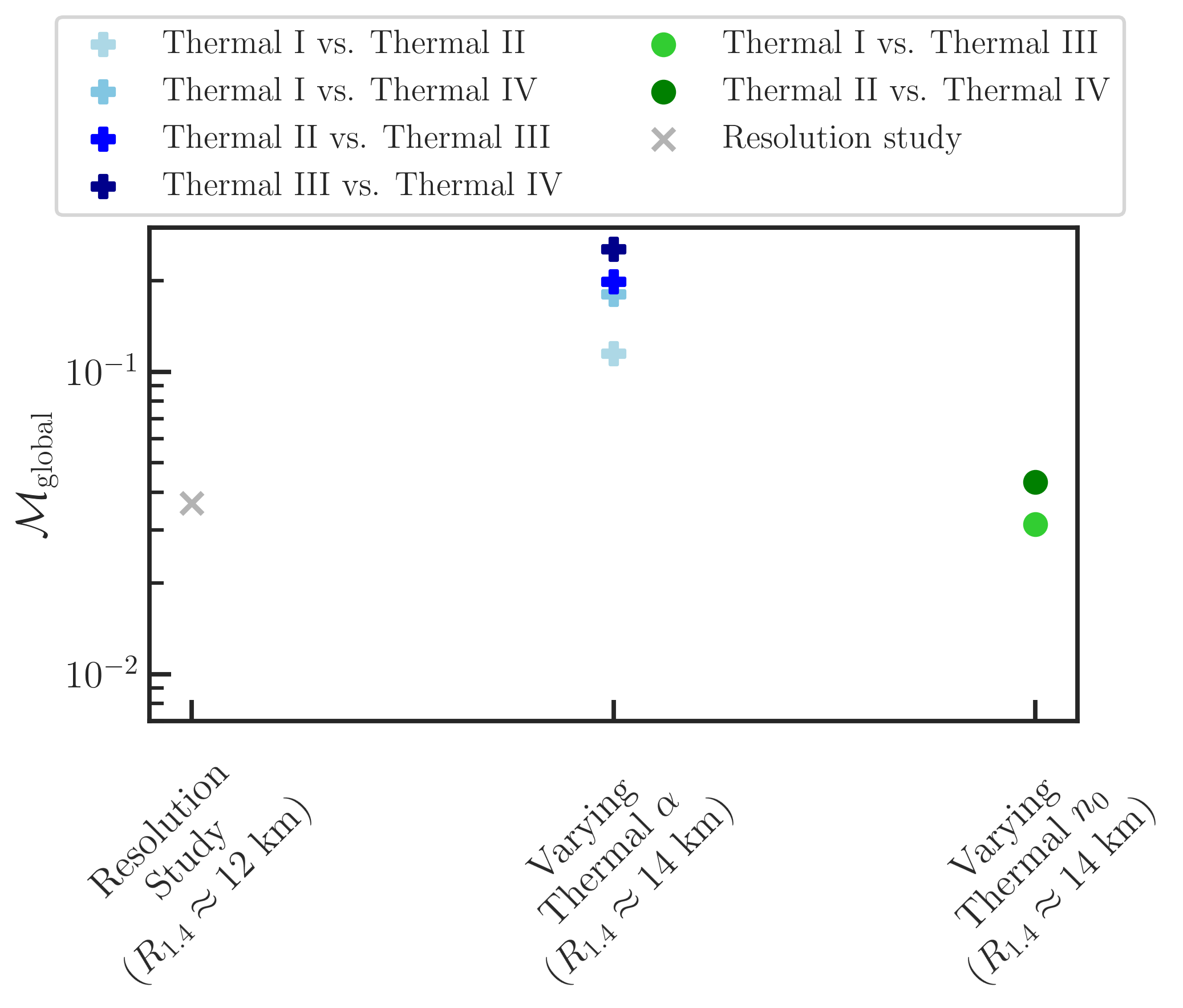}
\caption{\label{fig:peaks_R14} Same as Fig.~\ref{fig:peaks}, but for the set of \Rbig EoSs that vary in their thermal component. The resolution study (for an $R_{1.4}\approx 12$~km EoS) is repeated, for reference.}
\end{figure}

 In order to further understand the sensitivity of the spectra to changes in the underlying EoS,
the top panel of 
 Figs.~\ref{fig:peaks} and \ref{fig:peaks_R14} show the mismatches 
 calculated for the first two peaks individually, which are
 the most observationally relevant features, given that the third peaks occur at  higher 
frequencies which are unlikely to be resolved even with XG detectors. 
We define the mismatch of Peak 1  ($\mathcal{M}_1$; shown visually as the 
red shaded regions in Figs.~\ref{fig:spectra} and \ref{fig:spectra_R14}) via eqs.~\ref{eq:M}-\ref{eq:innerP},
with frequency bounds between $f_{\rm min}$ and the average of the first minima
of the two spectra being considered. The mismatch of Peak 2 ($\mathcal{M}_2$; corresponding
to the blue shaded regions in Figs.~\ref{fig:spectra} and \ref{fig:spectra_R14}) is likewise calculated between
the average first minima and the average second minima of the two spectra being compared.

As was seen visually in Fig.~\ref{fig:spectra} for
the case of the \Rsmall models, Fig.~\ref{fig:peaks} further demonstrates
that the $\Delta R_{1.4}=-120$ and $-54$~m cases
are characterized by relatively large mismatches in their first peaks. All of the cases
for which the thermal treatment is varied have smaller
mismatches in their first peaks.
In other words, varying the cold EoS towards smaller radii has a significant effect
on not only the dominant spectral peak, but also the lower-frequency first peak.
In contrast, varying the thermal prescriptions seems to primarily affect the
second (dominant) spectral peak. This has significant
implications for the detectability of these spectral differences, as we discuss further below.

The bottom panel of Fig.~\ref{fig:peaks} shows the global mismatches for
the same models,
calculated across the entire post-merger frequency range from $f_{\rm min}$=2~kHz
to $f_{\rm max}$=5.5~kHz. We see here that changing the $\alpha$ 
parameter of the thermal prescription (which governs the power-law decay of the effective
mass with density) leads to global mismatches that are comparable to changing the
cold EoS by -54 to +116~m, for all of the \Rsmall EoSs considered.

Figure~\ref{fig:peaks} also includes, as an approximate reference point, the mismatch from the simulations performed with two different resolutions, the higher of which matches the effective resolutions of the \Rsmall and \Rbig evolutions (see Appendix~\ref{sec:appGW} for details).
This reference mismatch confirms that the thermal differences are negligible
for EoSs that differ only in the value of the $M^*$ density-transition parameter, 
$n_0$, (shown in green). However, the mismatches in at least one of the spectral peaks
are indeed numerically significant for all other pairs of EoSs considered.

We quantify the relative mismatch between the first and second spectral peaks for the \Rbig EoSs in Fig.~\ref{fig:peaks_R14}, calculated using the same method described above but with appropriately adjusted frequency ranges. Although the first spectral peaks have slightly larger mismatches for these stiffer EoSs, 
we find that the second peak still dominates the overall mismatch by a factor of $\mathcal{M}_2/\mathcal{M}_1 \sim$ 9-16, for the models that vary in the thermal $\alpha$ parameters (blue crosses). In other words,
the effect of varying the thermal treatment for the \Rbig EoS is still primarily concentrated around the second spectral peak.

 From these observations, we draw a few key conclusions. First, the peak frequency
 alone is insufficient to encapsulate the entirety of the differences between two spectra, even
 in the limit of very small changes to the underlying EoS. This is illustrated most directly by
 the $\Delta R_{1.4}=-54$~m case, for which the spectrum differs significantly from the
 baseline model, despite negligible differences in $f_2$. Figure~\ref{fig:spectra} also shows that
 reducing $R_{1.4}$ by $-120$~m
 leads to a comparable shift in the peak frequency as increasing $R_{1.4}$ by $+116$~m,
 but the mismatch is significantly larger in the former case.
 These observations highlight the need 
 to consider global mismatches, together with the peak frequency, to fully characterize the differences between
 two spectra, and suggests some limitations to how completely the quasi-universal relations (which focus 
 on a single peak frequency) can be used to characterize a spectrum.
   
 Second, we find that varying the thermal treatment
 can lead to larger shifts in $f_2$ than varying the cold EoS by $\pm120$~m.
 Moreover, when considering the integrated (global) mismatch between spectra,
 we find that current uncertainties in the thermal physics lead
 to mismatches comparable to the mismatches that 
 are found when the cold EoS is varied by -54 to +116~m. This implies
 that if the cold EoS is known at the level of $\pm \sim120$~m in $R_{1.4}$
   -- as will be possible
 within $\sim$1~year of binary neutron star inspiral observations with CE \cite{Chatziioannou:2021tdi,Finstad:2022oni} --  
the post-merger GW spectra can be used to directly probe
 the finite-temperature part of the EoS.

\subsection{Distinguishability of the post-merger GWs}
We turn now to the question of how well these thermal signatures in the post-merger GWs
can be measured, with current and upcoming detectors. To quantify this,
we use the horizon distance, which defines the maximum distance
at which two signals can be distinguished according to
\begin{equation}
\label{eq:dhor}
    d_{\rm hor} = \frac{d_i \rho_i}{\rho_{\rm distinguish}}
\end{equation}
where $\rho_i$ is a reference signal-to-noise ratio (SNR)
evaluated at a distance $d_i$.  
We assume that the SNR of the individual spectra, $\rho_i$, are
similar to one another for a given pair of EoSs, and thus use their average SNR as the reference
when computing the horizon distances.
Finally, $\rho_{\rm distinguish}$ represents the
SNR required to distinguish between two GW spectra,
which is given by
\begin{equation}
\label{eq:distinguish}
\rho_{\rm distinguish}  = \max\left[ \frac{e}{\sqrt{2 \mathcal{M}}}, \rho_{\rm thresh} \right],
\end{equation}
for distinguishability at 90\% confidence  \cite{Baird:2012cu}. The threshold parameter,
$\rho_{\rm thresh}$, imposes a minimum SNR requirement
for detectability of the signal. We use a threshold for detectability
of  $\rm \rho_{\rm thresh}=5$.

\begin{figure}[!ht]
\centering
\includegraphics[width=0.45\textwidth]{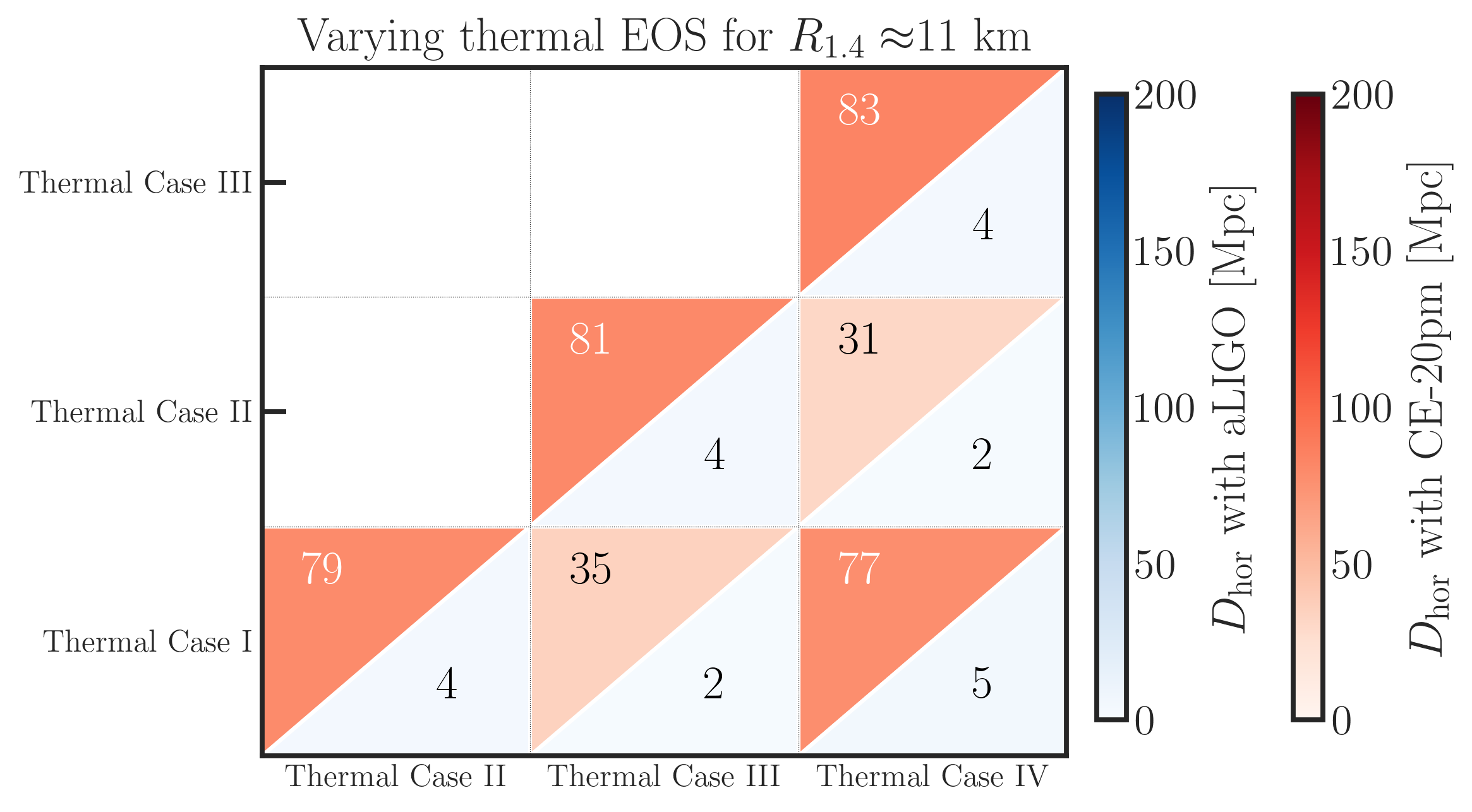} \\ 
\includegraphics[width=0.45\textwidth]{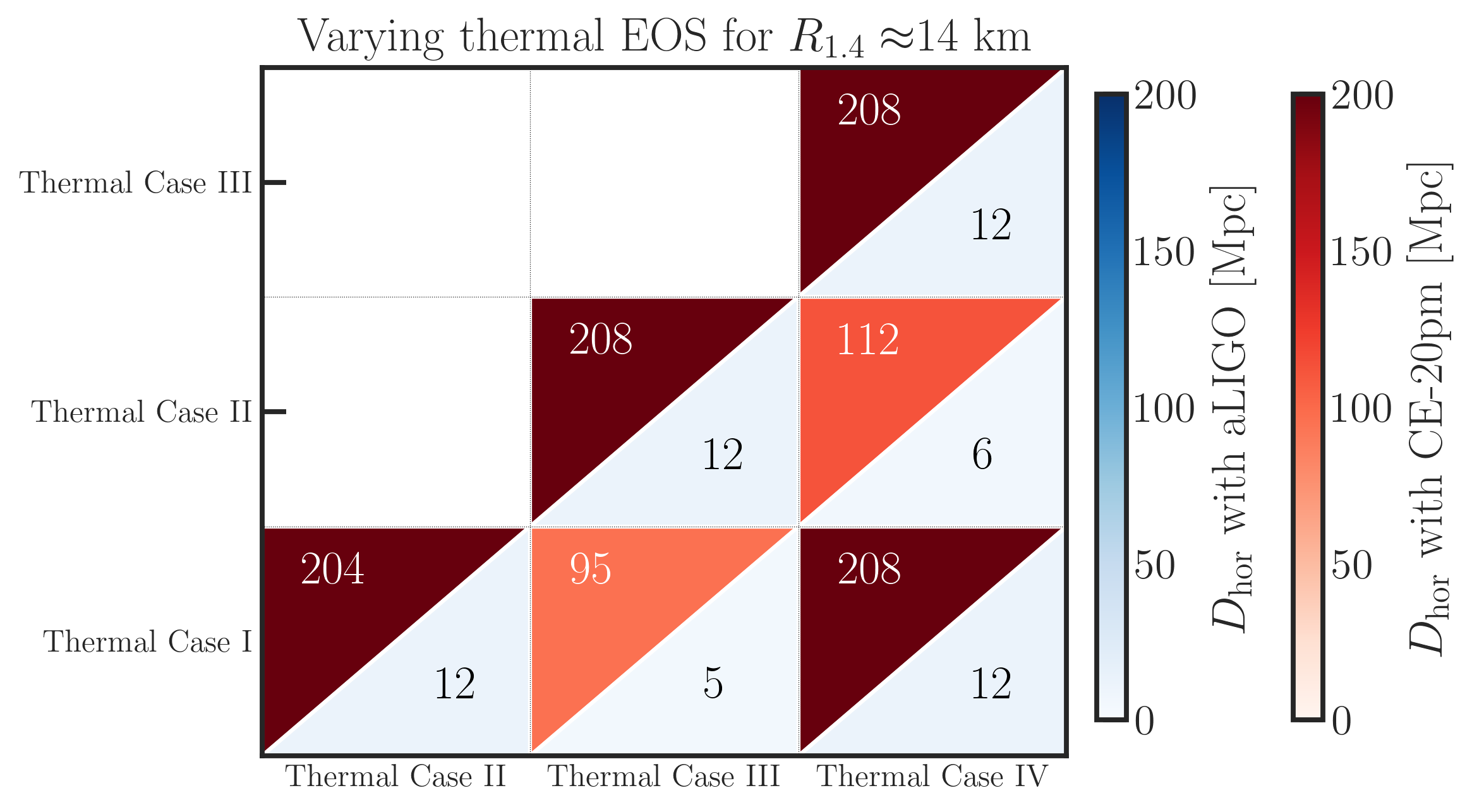} \\
\caption{\label{fig:dhor} Horizon distances (in Mpc) to which different pairs of EOS variations would be
distinguishable at 90\% confidence, assuming the design sensitivity of aLIGO (in blue) or the projected sensitivity of the CE-20pm
configuration (in red) and an SNR threshold for distinguishability of 5.}
\end{figure}

We use our three-peak templates (shown in Figs.~\ref{fig:spectra} and \ref{fig:spectra_R14}; see also Appendix~\ref{sec:appGW}) to compute the horizon distances.
We note that our  integration bounds of $f$=2-5.5~kHz (1.4-4~kHz) for the \Rsmall (\Rbig) EoSs 
intentionally exclude the inspiral contributions to the spectra, in order
to focus on the new information that can be gleaned post-merger. However, this choice
 also excludes lower-frequency differences that may emerge after the merger,
 such as from an $m=1$ one-arm instability 
 \cite{East:2015vix,Paschalidis:2015mla,Radice:2016gym}. Thus, the mismatches and horizon distances reported in this work
 should be viewed as conservative lower limits, as a more sophisticated template
 that includes such features might expose additional differences. 

Figure~\ref{fig:dhor} shows the resulting horizon distances for which various
pairs of EoSs could be distinguished. 
In the top panel, we show pairs of EoSs that vary in their thermal treatment with a soft cold EoS (\Rsmall),
 while in the bottom panel, we show the pairs of
models that vary in their thermal treatment with a stiff cold EoS (\Rbig).
We compute horizon distances
for two different detector noise curves: aLIGO shown in blue \cite{aLIGOsensitivity}, 
and the 20~km configuration
of Cosmic Explorer that has been tuned for post-merger sensitvity (CE-20pm) in red \cite{Srivastava:2022slt}.

With the design sensitivity of the current aLIGO instruments,
we find that, even for the most extreme differences in thermal treatment, the resulting
GW spectra would only be distinguishable for extremely nearby mergers (within 12~Mpc). 
Given that the closest neutron star merger every century is expected to be roughly
13$\substack{+9\\-4}$~Mpc away \cite{Burns:2019byj},
it not very likely that we will be able to directly measure
these thermal effects with the current generation
of detectors. However, with the sensitivity of CE-20pm, the horizon
distances increase significantly, to up to 83~Mpc and 208~Mpc for resolving thermal effects in the
\Rsmall and \Rbig cold EoSs, respectively.

It is interesting to observe that the thermal effects
are more easily distinguishable for the \Rbig EoSs,
despite the fact that the mergers evolved with
the \Rsmall cold EoS experience more heating 
and larger shifts in the peak frequencies (5\%
fractional variation in $f_2$ for the \Rsmall EoSs,
compared to 2\% shifts for the \Rbig EoSs; see Table~\ref{table:thermal}). This is driven
in part by the fact that the spectra peak at lower frequencies ($\sim2.6$~kHz) for 
the \Rbig evolutions, where the detectors
are significantly more sensitive. However,
the difference in the central peak frequency does not
entirely explain the larger mismatches. 

We confirm this with a simple test that shifts the CE-20pm noise curve down by 800~Hz, which
is the approximate difference in peak frequencies between the \Rsmall
and \Rbig EoSs. This artificially ensures that the \Rbig spectra peak at frequencies
where the detector is comparably sensitive. Even in this case, we find that
the horizon distances reach as far as $\sim$150~Mpc for the most disparate thermal treatments.
This implies that the spectra are significantly more distinguishable than we find
for the \Rsmall spectra, despite having smaller differences in $f_2$.
This is due to slightly larger differences in the first and third spectral speaks
for these EoSs;
although the effect of changing the thermal treatment is still most concentrated in
the second spectral peak (see Fig.~\ref{fig:peaks_R14}).

We note that, even though the thermal effects led to comparable
\textit{global} mismatches compared to the changes in the cold EoS
(see Fig.~\ref{fig:peaks}), the thermal
differences will be slightly harder to distinguish, than would
be estimated based on the ``equivalent" change to $R_{1.4}$.
As an example, for the baseline vs. $\Delta R_{1.4}=+116$~m cold EoS
comparison, the post-merger spectra could be distinguished at distances
of up to 107~Mpc with the sensitivity of CE-20pm.
The (slightly) lower horizon distances for the \Rsmall EoSs
with varied thermal prescriptions are a consequence of the frequencies
at which the spectra differ:  as was seen in Fig.~\ref{fig:peaks},
 varying the
cold EoS also produced significant differences in the first (lower-frequency)
peak, while the variations in the thermal treatment lead to
differences that are concentrated around the second peak. Although the
second peak is louder, it is at higher frequencies, where the
detectors are less sensitive.  This poses an additional
challenge for measuring thermal effects in the post-merger GWs; however,
the horizon distances found here indicate that such constraints are nevertheless
 within reach for the next generation of detectors.

\begin{figure}[!ht]
\centering
\includegraphics[width=0.45\textwidth]{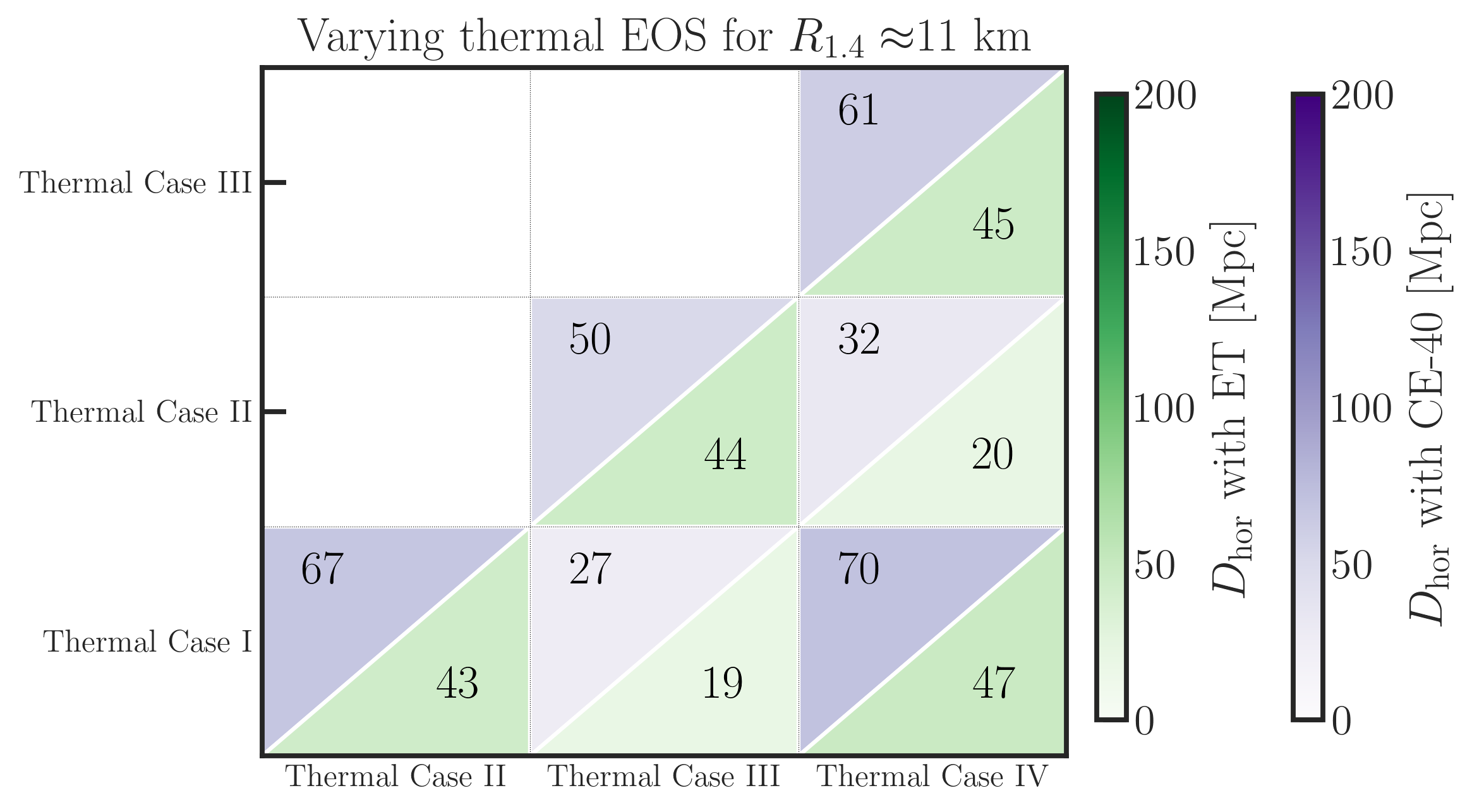}
\includegraphics[width=0.45\textwidth]{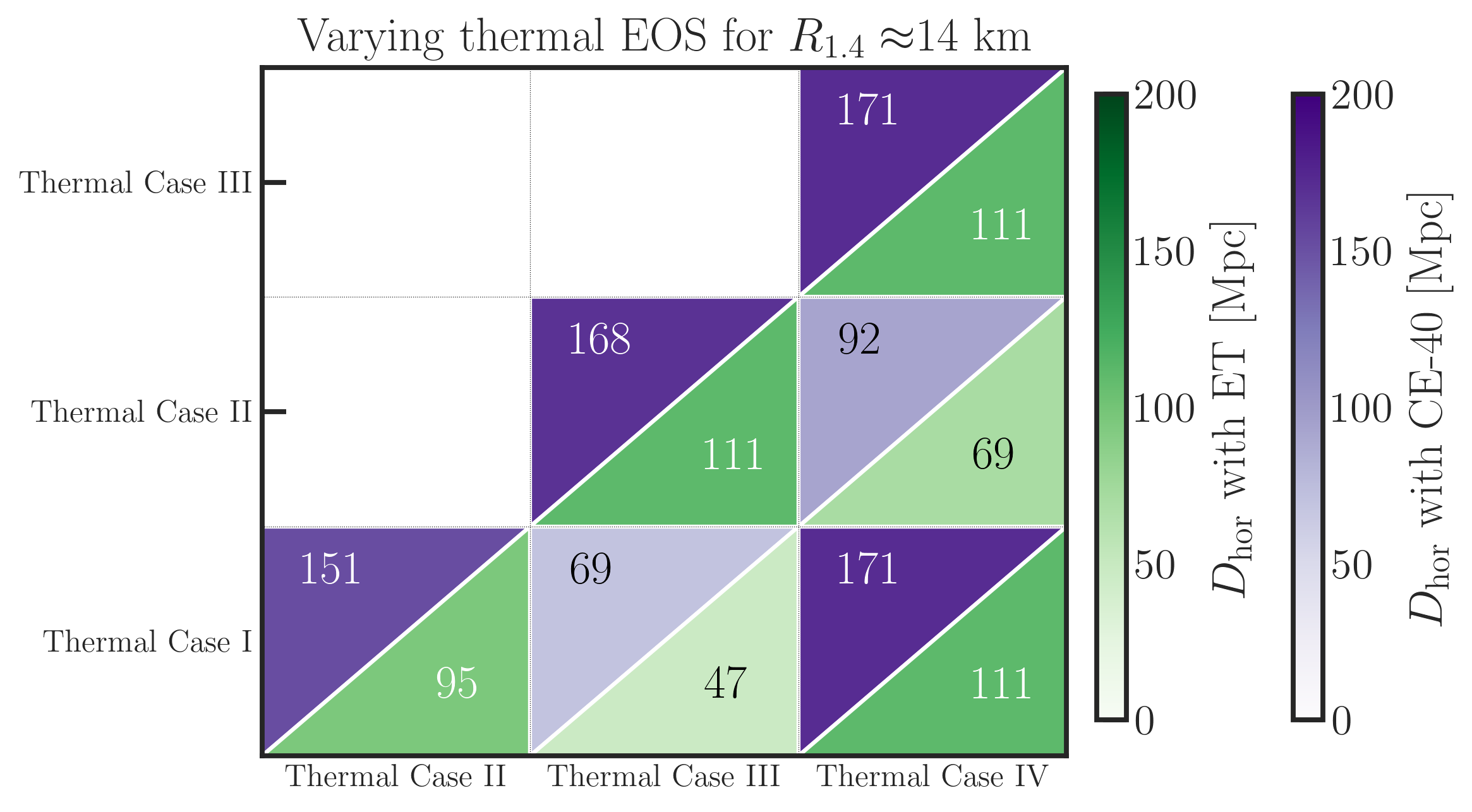}
\caption{\label{fig:ETCE40}  Horizon distances (in Mpc) to which different pairs of EOS variations would be
distinguishable at 90\% confidence, assuming the projected sensitivity of Einstein Telescope (in green) or the 40~km configuration of Cosmic Explorer
 (in purple) and an SNR threshold for distinguishability of 5. }
\end{figure}

Finally, we also compute the horizon distances for the same pairs of EoSs, assuming
the projected sensitivity of Einstein Telescope (ET) \cite{Punturo:2010zz, ETsensitivity} and the 40~km configuration
of Cosmic Explorer (CE-40) \cite{Reitze:2019iox,CEsensitivity} in Fig.~\ref{fig:ETCE40}. For distinguishing the thermal effects,
we find horizon distances of up to 70~Mpc (47 Mpc) for the \Rsmall EoSs and 171 Mpc (111 Mpc)
for the \Rbig EoSs, with CE-40 (ET). For both detectors, we find that at least some of the thermal prescriptions would be distinguishable for an
event at the distance of GW170817.
Thus, we conclude that the 20~km ``post-merger" tuned configuration of
CE gives the best prospects for resolving thermal effects, but that the XG era is highly promising
with any of these detectors.

\section{Discussion and Conclusions}
In this work, we have shown that the uncertainties in the finite-temperature
part of the EoS can alter the post-merger GWs by a degree comparable to 
changing the cold EoS by $\pm \sim 120$~m in radius (or, equivalently, by $\pm5$\% in the cold pressure
at supranuclear densities). This suggests that it may be possible
to derive new constraints on the physics of finite-temperature dense matter,
if the cold EoS can be pinned down at this level. Recent estimates
have suggested that observations of binary neutron star inspirals in the XG era
will constrain the cold EoS to within 50-200~m in radius \cite{Chatziioannou:2021tdi,Finstad:2022oni},
implying that such constraints will indeed be within reach.

We find that the finite-temperature effects lead to spectral differences that are dominated by
the second (dominant) peak, whereas varying the cold EoS towards smaller radii leads to
more significant changes in the first peak as well. Because the detector sensitivity
decreases rapidly with frequency, this makes the thermal effects slightly more challenging
to detect. Nevertheless, we find horizon distances of up to 83~Mpc for distinguishing
different thermal treatments for a soft cold EoS (\Rsmall) and up to 208~Mpc
for a stiffer cold EoS (\Rbig), when assuming the sensitivity of the 20~km post-merger
tuned configuration of CE. The maximum horizon distances for distinguishing thermal effects given
the 40~km configuration of CE
or ET are $\sim$20-85\% closer; but, in both cases, at least some of the thermal prescriptions
would still be distinguishable for an event at the distance of GW170817. 
Building a framework to measure thermal parameters from
a future GW signal requires additional parameter estimation studies,
beyond the scope of this work. Nevertheless, these results indicate
that the XG era is highly promising for distinguishing new
finite-temperature effects from the post-merger GWs with any of these proposed detectors,
with the CE-20pm configuration offering the best prospects, 
and they provide new motivation for parameter estimation studies to
further explore these detection prospects.

Finally, we note that the strength of the thermal effects is likely to
also depend on the total mass and mass ratio of the system, and may be
sensitive also to the treatment of neutrino transport
\cite[e.g.,][]{Fields:2023bhs}. Other effects such as non-zero initial
spins \cite{East:2019lbk}, bulk viscosity arising after the merger
\cite{Hammond:2022uua,Most:2022yhe}, or the effects of including
additional degrees of freedom
\cite[e.g.,][]{Sekiguchi:2011mc,Radice:2016rys,Bauswein:2018bma,Most:2019onn,Blacker:2023opp,Espino:2023llj,Vijayan:2023qrt}
may also contribute to spectral distortions of similar magnitude to
what have been studied here. In particular, some classes
of phase transitions may not be constrained by
the inspiral GWs alone, posing an additional challenge for pinning
down the cold EoS at the level required for constraining
these thermal models \cite{Raithel:2022efm,Raithel:2022aee}.
Finally, while our templates accurately
capture the features of the three primary spectral peaks, they may
miss out on additional differences between the spectra such as the
emergence of spiral modes or one-arm instabilities.  By construction, our smoothed templates also do not capture the finest-scale features of the spectra.
As a result, the
horizon distance estimates provided in this work should not be
considered as the final word. Rather, these horizons provide a
promising indication that differences in the thermal physics can be
measurable with next-generation facilities, if the cold EoS is
sufficiently well constrained, and motivate the construction of
faithful post-merger GW templates, to fully
characterize the possible interplay with these additional effects.

\begin{acknowledgments}
CR is supported by a joint postdoctoral
fellowship at the Princeton Gravity Initiative
and the Institute for Advanced Study, with support from the John N. Bahcall Fellowship Fund and Schmidt Futures. This
work was in part supported by NSF Grant PHY-2145421 to the University of Arizona. 
The simulations presented in this work were carried out with
the Stampede2 cluster at the Texas Advanced Computing Center
and the Expanse cluster at San Diego Supercomputer Center, under XSEDE allocation PHY190020.
The simulations were also performed, in part, with the
Princeton Research Computing resources at Princeton
University, which is a consortium of groups led by the
Princeton Institute for Computational Science and Engineering 
(PIC-SciE) and Office of Information Technology’s Research Computing
\end{acknowledgments}

\appendix

\section{Construction of the cold equations of state}
\label{sec:appEOS}

In this appendix, we describe the construction of the sequence of cold EoSs
used in this work. For the baseline \Rsmall model and the \Rbig models, we use
a generalized piecewise-polytropic (PWP) parametrization  
of the wff2 \cite{Wiringa:1988tp} or H4 EoSs \cite{Lackey:2005tk} at high
densities, with a parametrized representation of SLy \cite{Douchin:2001sv} for the crust. 
We follow the generalized PWP framework of \cite{OBoyle:2020qvf} 
and parametrize the EoS via three segments, which 
are divided at the fiducial densities $\rho_{1}=10^{14.87}$~g/cm$^3$ and
$\rho_{2}=10^{14.99}$~g/cm$^3$. The pressure along a given segment is given by
\begin{equation}
P(\rho) = K_i \rho^{\Gamma_i} + \Lambda_i, \quad	\rho_{i-1} < \rho \le \rho_i,
\end{equation}
where the coefficient, $K_i$, is determined by requiring differentiability,
\begin{equation}
\label{eq:Ki}
K_i = K_{i-1} \left( \frac{\Gamma_{i-1}}{\Gamma_i} \right) \rho_{i-1}^{\Gamma_{i-1}-\Gamma_i},
\end{equation}
and the parameter $\Lambda_i$ is imposed to ensure continuity, such that
\begin{equation}
\label{eq:Lambdai}
\Lambda_i = \Lambda{i-1} + \left( 1- \frac{\Gamma_{i-1}}{\Gamma_i} \right) K_{i-1} \rho_{i-1}^{\Gamma_{i-1}} .
\end{equation}
There are thus four free parameters: 
${K_1, \Gamma_1, \Gamma_2, \Gamma_3}$. From these parameters and the low-density
EoS, all other $K_{i}$ and $\Lambda_i$ are uniquely determined.
The fit procedure to construct the \Rsmall and \Rbig EoSs has been described 
previously in Appendix A of \cite{Raithel:2023zml}, but we 
repeat the relevant model parameters in Table~\ref{table:GPP} for convenience.

  \begin{table}
  \centering
\begin{tabular}{cccccc }
\hline 
$R_{1.4}$ [km]  & $\rho_0$ [g/cm$^3$] &  $\log_{10} K_1$  &   $\Gamma_1$   &  $\Gamma_2$ &  $\Gamma_3$   \\
\hline \hline 
 11.12 & 1.309 $\times 10^{14}$ &  -35.443 &  3.316 &  4.122 &  3.200   \\
 13.99 & 2.931 $\times 10^{13}$ & -23.110 &  2.502 &  1.511 &  2.366  \\
\hline
\end{tabular}
\caption{\label{table:GPP} Model parameters for the generalized piecewise polytropic
representations of the \Rsmall and \Rbig EoSs. 
$R_{1.4}$ indicates the radius of a 1.4~$\Ms$ neutron star predicted by each EoS.
The parameter $\rho_0$ is the density at which the high-density
parametrization intersects the crust EoS, which is taken to be a GPP
 representation of SLy. The remaining four columns
 provide the four free parameters used to characterize eqs. \ref{eq:Ki}-\ref{eq:Lambdai}.
 Table is repeated from \cite{Raithel:2023zml} with permission. }
 \end{table}

\begin{figure}[!ht]
\centering
\includegraphics[width=0.45\textwidth]{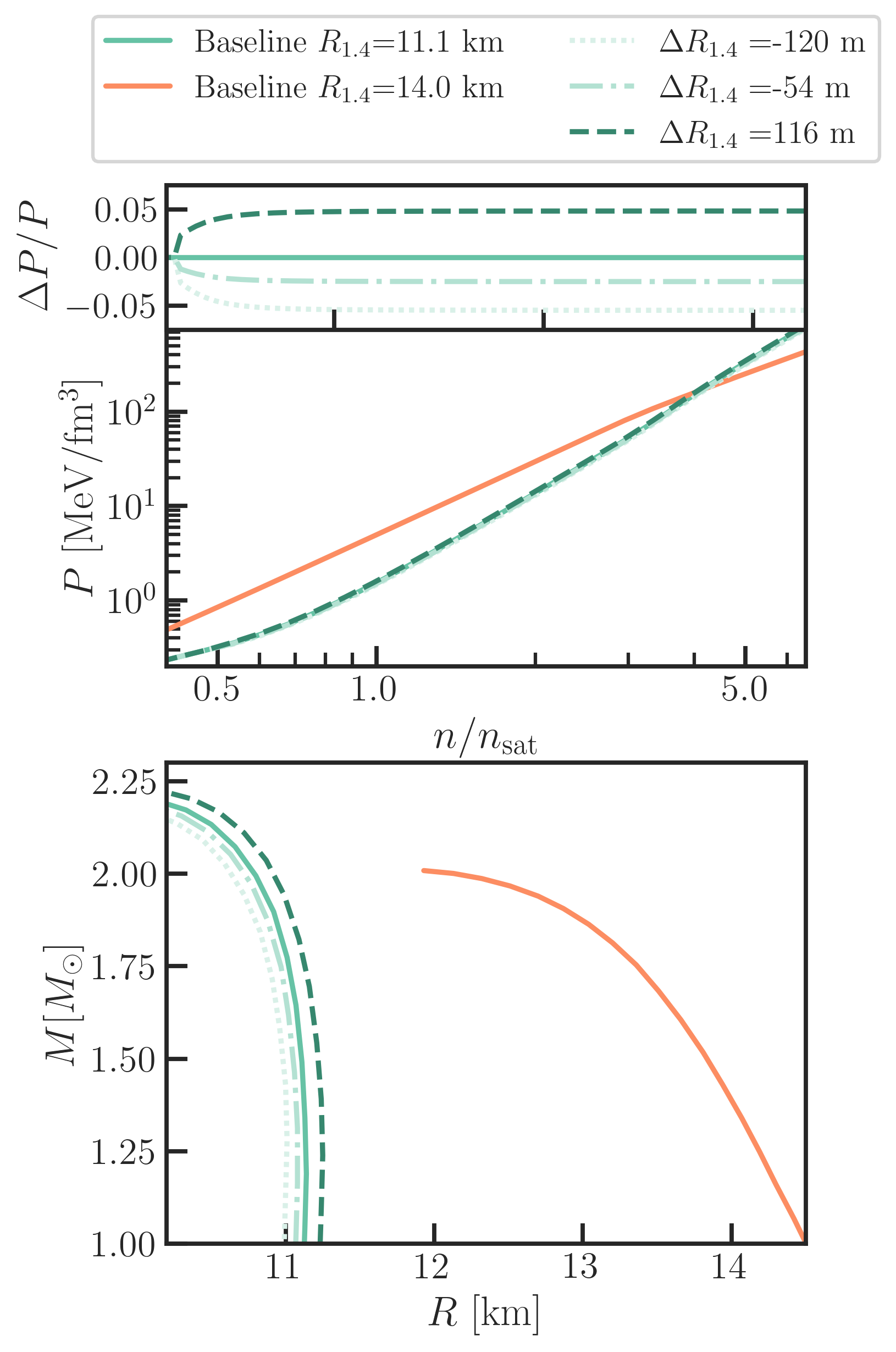}
\caption{\label{fig:eos} Cold equations of state considered in this work. 
Top - inset: fractional difference in the zero-temperature
pressure for the $R_{1.4}\approx11$~km models, compared to the baseline model. 
Top - main figure: Pressure as a function of the density, relative to
the nuclear saturation density $n_{\rm sat}$=0.16~fm$^{-3}$. Bottom: corresponding mass-radius relations.}
\end{figure}

To construct a sequence of new models that deviate by a small degree from the baseline \Rsmall EoS,
we systematically scale the pressure above $\rho\ge 1.2\rns$ by a constant
factor, ranging from 0.95 to 1.05. Between $\rns$ and $1.2 \rns$, we linearly connect between
this scaled pressure and the baseline model, which we use for all densities below $\rns$.
The resulting EoSs are shown in Fig.~\ref{fig:eos}, and their characteristic properties
are summarized in Table~\ref{table:coldEOS}.

  \begin{table}
  \centering
\begin{tabular}{cccc }
\hline 
EoS  & $P_c$ scale factor &  $R_{1.4}$ [km]  & $\Lambda_{1.4}$    \\
\hline \hline 
$\Delta R_{1.4}=-120$~m & 0.95 & 11.00 &  214   \\
$\Delta R_{1.4}=-54$~m & 0.975 & 11.07 &  223   \\
Baseline & 1.0 & 11.12 &  230   \\
$\Delta R_{1.4}=+116$~m & 1.05 & 11.24 &  244   \\
\hline
\end{tabular}
\caption{\label{table:coldEOS} Characteristic properties of the sequence of cold EoSs that deviate
by a small degree from the \Rsmall baseline model. The first column indicates
the labels used throughout this work. The second column lists the factor by which
the pressure is scaled at all densities above 1.2$\times$ the nuclear saturation density,
compared to the baseline model. The final two columns report the characteristic radius
and tidal deformability for a 1.4~$\Ms$ neutron star.}
\end{table}

In extending these cold EoSs to finite-temperatures, we follow the framework
of \cite{Raithel:2019gws}, with all details identical to the implementation in \cite{Raithel:2023zml}.

\section{Temporal evolution of the post-merger gravitational waves}
\label{sec:appSpec}

\begin{figure}[!ht]
\centering
\includegraphics[width=0.45\textwidth]{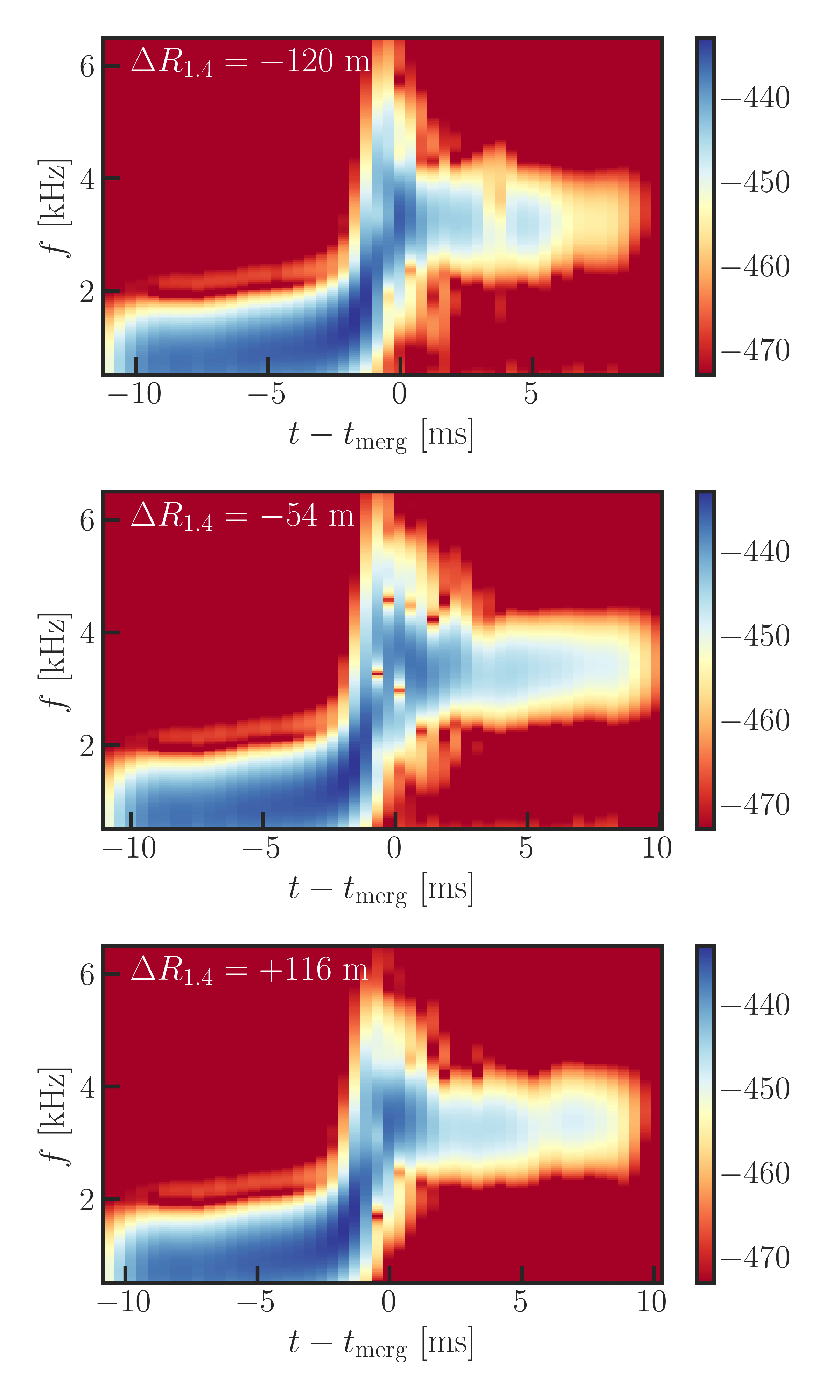}
\caption{\label{fig:sgram_cold} Spectra for the
\Rsmall cold EoSs. The strain includes the dominant $\ell=m=2$ 
mode, and assumes a face-on orientation for a source directly overhead at a distance of 40 Mpc. The power is shown by the color, with decibel ($20 \log_{10}$) scaling.  }
\end{figure}

\begin{figure*}[!ht]
\centering
\includegraphics[width=0.8\textwidth]{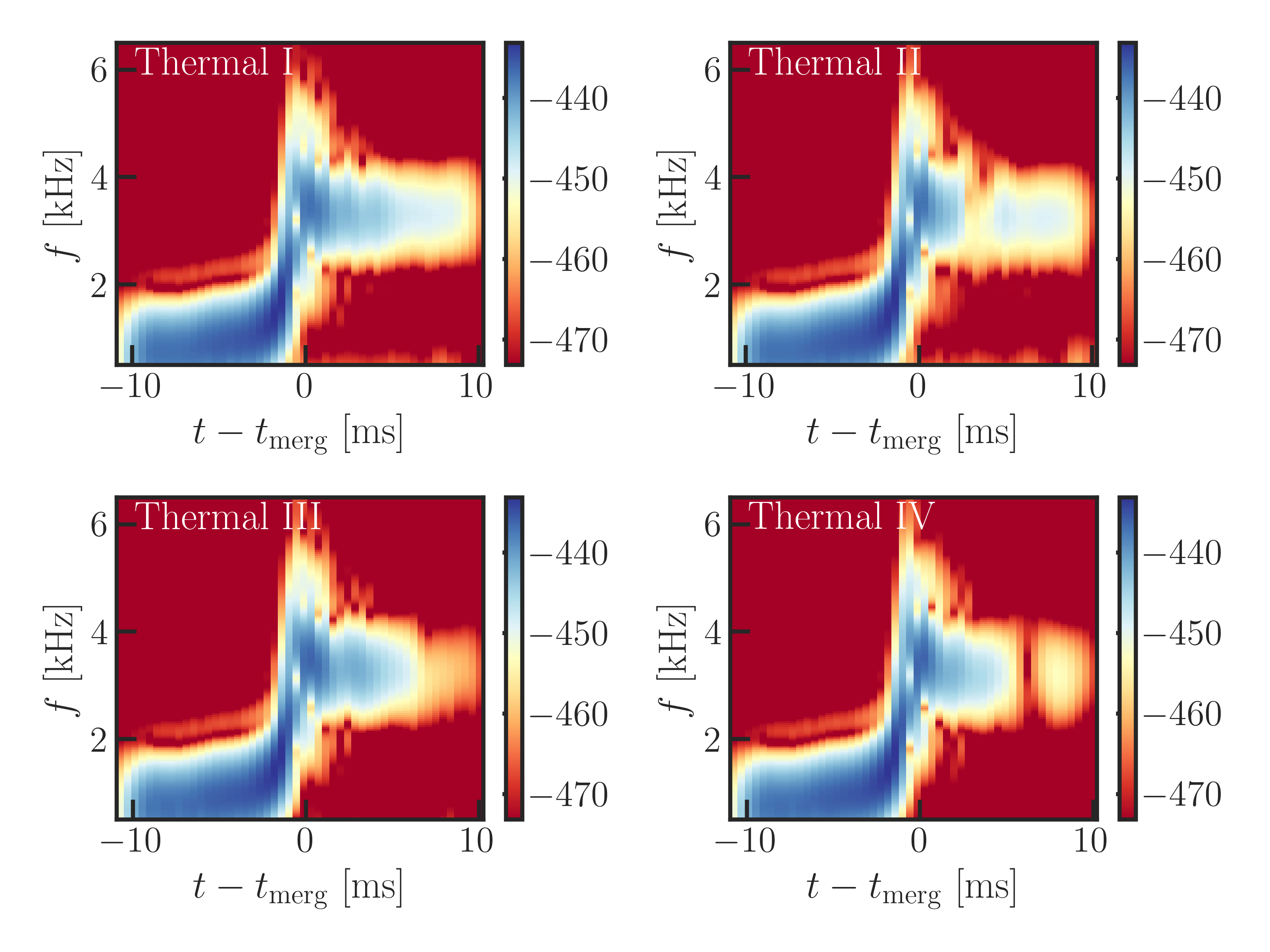}
\caption{\label{fig:sgram_R11} Same as Fig.~\ref{fig:sgram_cold}, 
but for the \Rsmall EoSs with different thermal treatments.}  
\end{figure*}

\begin{figure*}[!ht]
\centering
\includegraphics[width=0.8\textwidth]{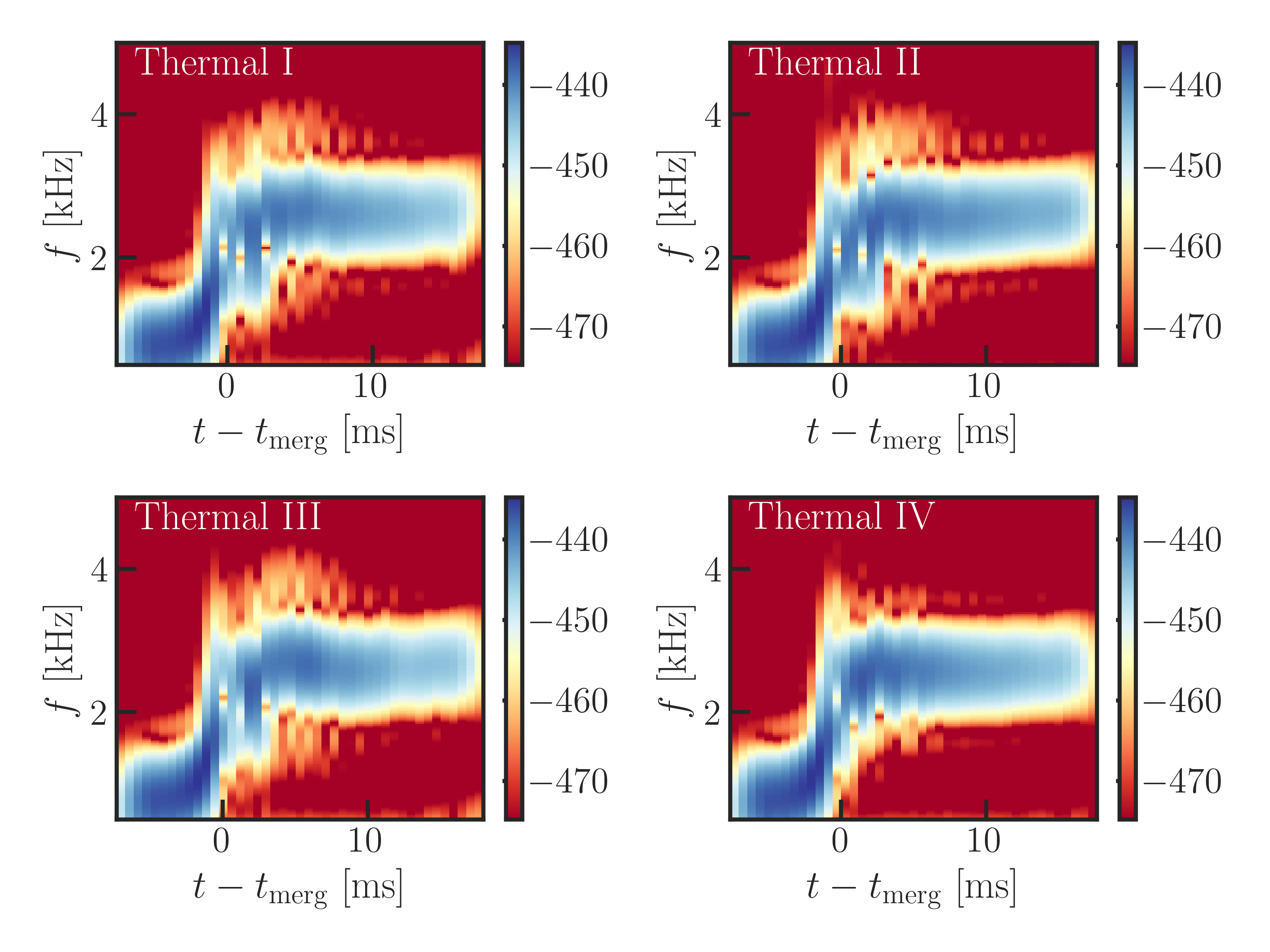}
\caption{\label{fig:sgram_R14} Same as Fig.~\ref{fig:sgram_cold}, 
but for the \Rbig EoSs with different thermal treatments.}  
\end{figure*}

In this appendix, we analyze the temporal evolution
of the post-merger gravitational wave signals for our
simulations. To that end, 
we compute spectrograms of the strain, $h(t)$,
using time bins of length 350$\Ms$, with overlap of 75\%. 
We show these spectrograms in
Figs.~\ref{fig:sgram_cold}-\ref{fig:sgram_R14}.

The GW power for the \Rsmall models decays
relatively quickly, within the first $\sim 10$~ms 
post-merger. In contrast, the \Rbig spectra are
more temporally extended, with significant
power still present at $10$~ms after the merger. 
We also note the presence of a small drift in the peak
frequencies for some models with the \Rbig cold
EoS  (e.g., thermal prescription II; top right figure of
Fig.~\ref{fig:sgram_R14}). In order to capture 
these late-time features, we evolved the \Rbig 
models to $\sim$20~ms after the merger, at
which time the strain is approximately stationary
and the power starts decaying.

\section{Post-merger gravitational waves and template fits}
\label{sec:appGW}
We extract the post-merger gravitational waves (GWs) from our simulations using
the Newman-Penrose scalar, $\psi_4$, which is decomposed onto $s=-2$ spin-weighted
spherical harmonics at large radii ($r=300 \Ms$). We calculate the  $+$ and $\times$ polarizations
of the GW strain via the relation, $\psi_4 = \ddot{h}_{+} - i \ddot{h}_{\times}$, using the fixed-frequency
integration method of \cite{Reisswig:2010di}. 

\begin{figure}[!ht]
\centering
\includegraphics[width=0.45\textwidth]{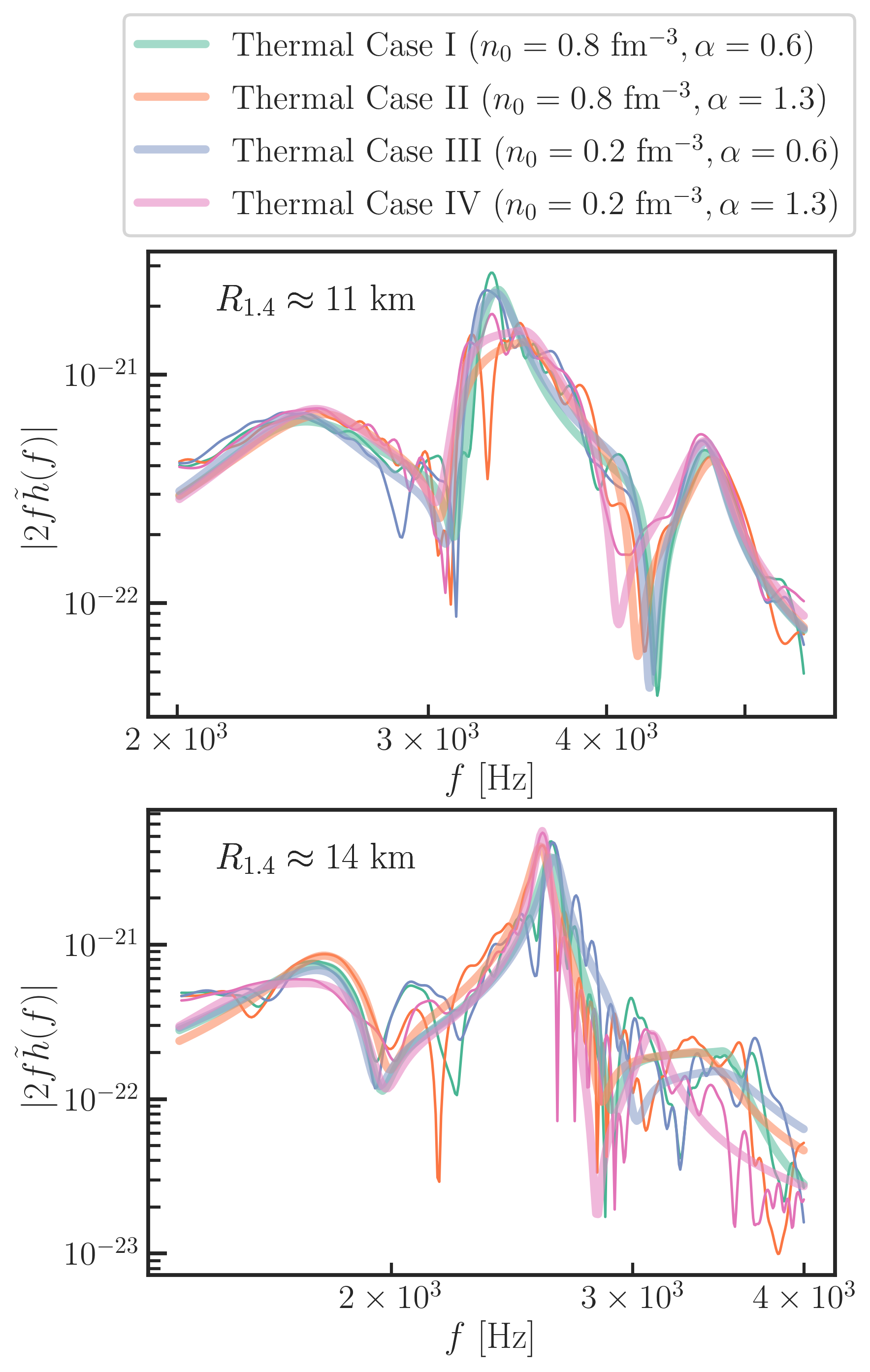}
\caption{\label{fig:all_spectra} Best-fit templates for the post-merger GW spectra. The \Rsmall
EoSs with different thermal treatments are shown in the top panel, while the \Rbig EoSs
are shown in the bottom panel. Templates (in thick, smooth lines) are fit to the raw spectra (thin, darker lines) 
including the dominant $\ell=m=2$ mode,
assuming a face-on orientation with an optimal detector response
factor and a source distance of 40~Mpc. }
\end{figure}

To compute the spectra shown in this work, we first apply a Tukey window with shape parameter of 0.25
to the time-series strain, and window the post-merger signal
to have the same length, for all simulations within
a given comparison (e.g., all \Rsmall models or all \Rbig models).
We then calculate the characteristic strain according to
\begin{equation}
h_c = 2 f |\tilde{h}(f)|,
\end{equation}
where $\tilde{h}(f)$ is the Fourier transform of the strain, $h(t) = {h}_{+} - i {h}_{\times}$.
Finally, we apply a fourth-order Butterworth bandpass filter to the characteristic strain, in order to excise the inspiral
contribution and any high-frequency noise. We apply the filter over a frequency range of 2-5.5~kHz for the \Rsmall EoSs
and 1.4-4~kHz for the \Rbig EoSs. These frequency bounds conservatively bracket the main post-merger signal,
which we take to start just below the instantaneous GW frequency at merger and to end when the signal has 
dropped $\sim100\times$ below the amplitude of the dominant spectral peak. 
In computing the spectra used in this work, we assume a face-on orientation and an optimal detector
response, corresponding to a signal directly overhead.
We show the resulting spectra, including the dominant $\ell=m=2$ mode, as the thin lines in Fig.~\ref{fig:all_spectra}.

These spectra exhibit a high degree of noise on small-frequency scales, which we find
leads to biases in the calculation of the mismatch integrals. Such high-frequency noise can be the result of either numerical error or stochasticity due to the highly turbulent state of the merger remnant following merger.
In order to reduce the dependence of
the mismatches on these small-scale features, which are not physically significant, for e.g., a signal-to-noise calculation,
we fit the spectra with a simplified template model. 
A variety of phenomenological templates have been introduced previously, ranging
from templates based on sine-Gaussian wavelets \cite[e.g.,][]{Chatziioannou:2017ixj,Wijngaarden:2022sah} to
exponentially damped sinusoids \cite[e.g.,][]{Hotokezaka:2013iia,Bauswein:2015vxa,Yang:2017xlf,Tsang:2019esi,Soultanis:2021oia}.
Other approaches have constructed templates directly from numerical relativity
data, e.g. using a reduced basis set derived from principal component analysis \cite{Clark:2015zxa}
or using a hierarchical model \cite{Easter:2018pqy}. Recently, \cite{Breschi:2022xnc}
introduced a ``partially-informed" template, that combines a wavelet basis set with
relations calibrated to a large database of numerical relativity
simulations.

In this work, our goal is to agnostically capture the three dominant spectral peaks present in 
our simulations. To that end, we adopt a model constructed from three Lorentzian profiles (similar to the approaches of 
\cite{Hotokezaka:2013iia,Bauswein:2015vxa,Yang:2017xlf,Tsang:2019esi,Soultanis:2021oia}), and 
we additionally allow for non-zero skew to capture possible asymmetries in the peaks, which
we find improves the quality of the fits. An
individual profile is given by
\begin{equation}
\tilde{h}_i(f) = A_i \left[ 1 + \left( \frac{x}{ 1 + k_i~ \mathrm{sgn}[x]} \right)^2 \right]^{-1}  (1 - i x)
\end{equation}
where $A_i$ is the amplitude of the peak, $k_i$ is a skew parameter that governs the peak asymmetry, the
convenience parameter $x$ is defined as
\begin{equation}
x = \frac{f-f_{\mathrm{peak},i}}{Q_i},
\end{equation}
and $Q_i$ is related to the width of the peak.

We stitch together the individual peaks using a hyperbolic tangent smoothing, according to
\begin{equation}
\tilde{h}(f) = \tilde{h}_1(f) + \chi_1 (1- \chi_2) \left[ \tilde{h}_2(f) - \tilde{h}_1(f) \right] + \chi_2 \left[ \tilde{h}_3(f) - \tilde{h}_1(f) \right], 
\end{equation}
where $\chi_{j=1,2}$ are the smoothing functions, which are given by
\begin{equation}
\chi_j = \frac{1 + \tanh[0.01 (f-f_{\mathrm{trans},j}) ]}{2}
\end{equation} 
and $f_{\mathrm{trans},j}$ is the transition frequency that divides the adjacent spectral peaks.

In total, there are thus 11 parameters, including $A_i, Q_i, f_{\mathrm{peak},i}$
 for each of the three peaks and $f_{\rm trans,j}$ for the two dividing frequencies.
 In order to simplify our fits, we fix the peak frequencies to the values
 obtained from Welch-averages of our spectra. This procedure
 averages overlapping segments of the strain when computing the Fourier transform,
 and helps to more robustly identify the spectral peaks by reducing the noise.
 We note that the peak frequencies in Table~I differ slightly from those reported in Paper I,
 due to differences in the Welch-averaging procedure. In Paper I,
 we required the \Rsmall and \Rbig to have the same number of overlapping segments
 in the Welch averages, while in the present work, we require the length
 of the overlapping segments to be the same length (4~ms) for all simulations.
 The differences between these two conventions are small and do not affect the main
 results of either work.
 
 After fixing the peak frequencies in this way, we are left with
 8 free parameters, which we fit to the (un-averaged) spectra. 
 We perform a least-squares minimization to fit the magnitude
 of the characteristic strain. Although fitting
 the magnitude leaves the phase
 unconstrained, the template is used 
 only to calculate overlaps, which are subsequently
 maximized with respect to the phase. We weight the fit by the
 projected noise curve for the CE-20pm detector 
 \cite{Srivastava:2022slt}. 
 We show
 the resulting best-fit templates together with the original spectra, in Fig.~\ref{fig:all_spectra}.

\begin{figure}[!ht]
\centering
\includegraphics[width=0.45\textwidth]{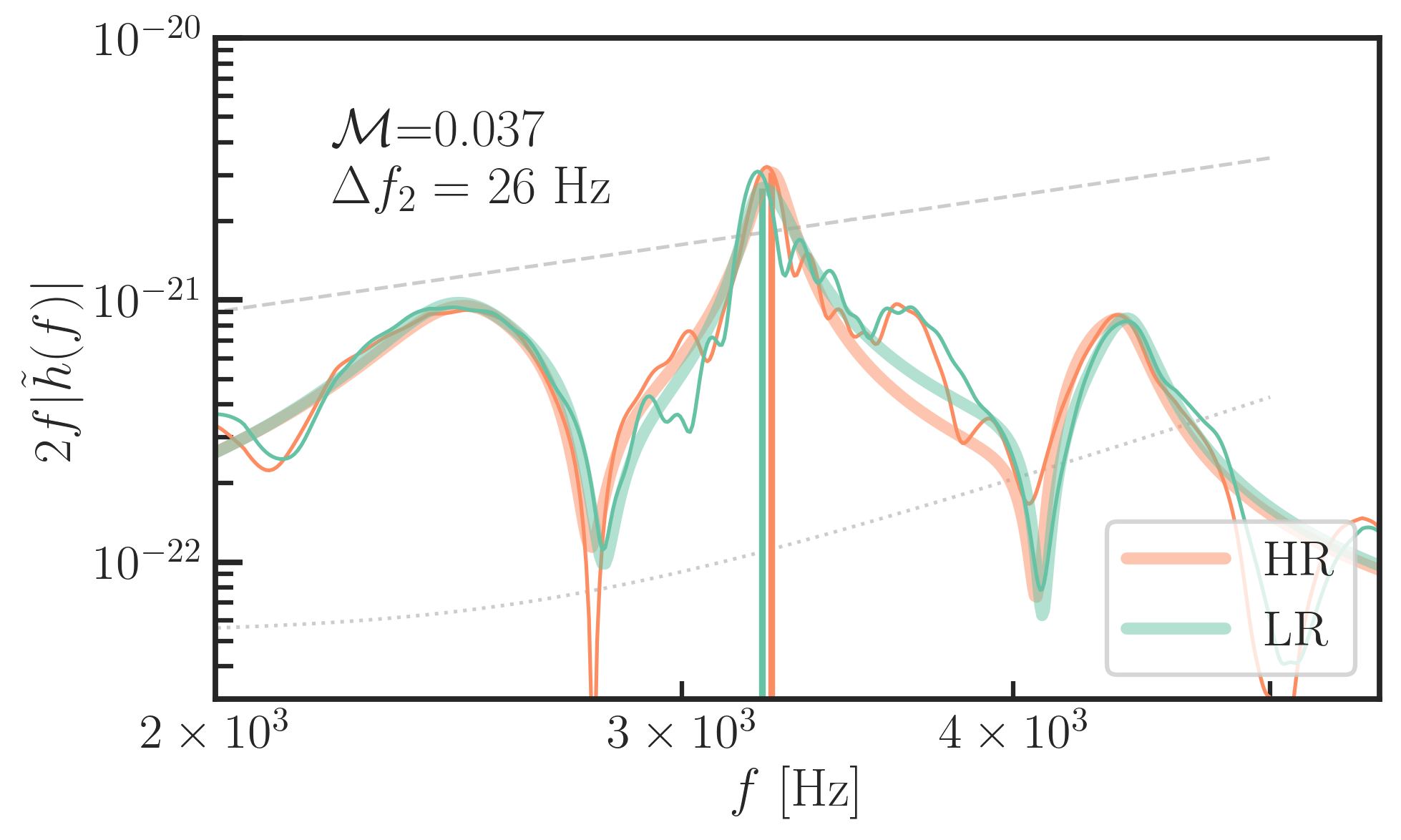}
\caption{\label{fig:multiRes} Post-merger GW spectra (thin lines) and best-fit templates
 (thick lines) for two different resolutions. The low resolution ($dx=195$~m) evolution is shown
 in teal, while a higher resolution ($dx=156.25$~m) evolution is shown in orange. The EoS corresponds
 to an intermediate stiffness with $R_{1.4}=12$~km \cite{Raithel:2022san}.}
\end{figure}

Figure~\ref{fig:multiRes} shows the sensitivity of the post-merger spectra and corresponding templates to
the resolution, for an EoS with intermediate stiffness ($R_{1.4}=12$~km). This EoS also uses an intermediate
set of $M^*$-parameters ($n_0=0.12$~fm$^{-3}, \alpha=0.8$), making it a convenient (albeit approximate) reference point
for the simulations explored in this work. The two resolutions correspond to a grid spacing on the
innermost refinement level of $dx=195$~m and $dx=156.25$~m; or, equivalently, to placing $\sim$100
or 125 points across the coordinate diameter of the initial neutron stars. The effective resolution of the high-resolution evolution is thus comparable to the resolutions used in the present work (i.e., covering the coordinate diameter of the initial neutron stars by $\sim$125 grid points; see Sec.~\ref{sec:IC}).
For further details on this EoS and the evolutions from which these
spectra were extracted, see \cite{Raithel:2022san}.
We find negligible changes to the spectra as the resolution is increased by a factor of 1.25. 
The dominant spectral peaks match identically and the mismatch between the best-fit spectra for the two 
resolutions is small as well, with $\mathcal{M}=0.016$. 

\begin{table}
  \centering
\begin{tabular}{ccccc }
\hline 
Cold EoS & \pbox{20cm}{\vspace{0.2cm}Variation to\\\vspace{-0.05cm}Cold EoS \vspace{0.2cm} }  & \pbox{20cm}{\vspace{0.2cm}Thermal \\\vspace{-0.05cm}Case \vspace{0.2cm}} &  $R^2$    \\
\hline \hline 
 &  $\Delta R_{1.4}=-120$~m 	       & I &  0.85  \\
\Rsmall &   $\Delta R_{1.4}=-54$~m  & I &  0.88  \\
&   $\Delta R_{1.4}=+116$~m	        & I &  0.89 \\
\hline
 	    && I &   0.91  \\
\Rsmall & -- & II &  0.88  \\
 	    && III &  0.94 \\
 	    && IV & 0.95  \\
\hline
 	    && I &  0.91 \\
\Rbig   & -- & II &  0.96   \\
 	    && III &  0.86 \\
 	    && IV & 0.96   \\
\hline
\end{tabular}
\caption{\label{table:R2}  Coefficient of determination, $R^2$, between the best-fit spectral templates
and the (un-averaged) characteristic strain from our simulations.}
\end{table}

We report the coefficients of determination for the template fits in
  Table~\ref{table:R2}.  Given that our fits provide signals that have approximately the same SNR as the raw signals they are based on, we compute the $R^2$ statistic as a measure of goodness-of-fit. We find that the best-fit templates account for 84-97\% of the variation
 in the underlying spectra. To put this number in context, we also calculate the
 $R^2$ statistic between the two different resolutions, using the complete spectrum (i.e.,
 not the template) from the low resolution simulation as an approximant of the
 high-resolution spectrum. We find that the low-resolution
 spectrum accounts for 89\% of the variation found in the high-resolution spectrum, 
 which is a comparable $R^2$ value to what we obtain for our template fits.
 In other words, the $\lesssim$15\% ``loss" we incur by using the templates in lieu of the
 complete spectra is comparable to the losses obtained by using a standard, instead of 
 high, resolution in our evolutions. 
 
We note that using the templates leads to mismatches that are always smaller than would be estimated from the raw spectra, because the templates effectively smooth out the smallest-scale differences. We report the mismatches calculated with
both the raw spectra and our best-fit templates, for all combinations of models, in Table~\ref{table:mismatches}. In all cases except one, the mismatches 
calculated with the templates are smaller than mismatches calculated with the
raw spectra. For the one exception to this statement (the \Rsmall model with Thermal Case III vs. IV), the mismatches are similar with the two calculation methods.
The smaller mismatches require closer source distances for the models 
to be distinguished (eq. ~\ref{eq:dhor}). In other words, using 
the smoothed spectral templates makes it harder to distinguish between the various
models.
For these reasons, we use the templates when estimating the distinguishability (and, e.g., calculating horizon distances) in order to be as conservative as possible.

\begin{table*}
\centering
\begin{tabular}{cccccc}
\hline \hline
Shared EoS    & Model A  &  Model B &  $\mathcal M_{\rm raw}$ &  $\mathcal M_{\rm template}$  & $\mathcal M_{\rm template}/\mathcal M_{\rm raw}$   \\
 \hline
     		  & Thermal I       	& Thermal II &  0.18 &  0.08  & 0.44\\
		 & Thermal I  		& Thermal III &  0.06   &  0.02 & 0.25 \\
\Rsmall ~~ & Thermal I            & Thermal IV  &  0.09   &  0.08 & 0.81  \\
		 & Thermal II           & Thermal III &  0.15 &  0.08  & 0.58 \\
		 & Thermal II  		& Thermal IV &  0.06   &  0.01  &  0.20 \\
     		& Thermal III  		& Thermal IV &  0.07 &  0.09  & 1.3 \\
\hline \hline
     		  & Thermal I       	& Thermal II &  0.48&  0.14  & 0.30 \\
		 & Thermal I  		& Thermal III &  0.08   &  0.03  & 0.40 \\
\Rbig ~~ & Thermal I            & Thermal IV  &  0.36  &  0.19  & 0.51 \\
		 & Thermal II           & Thermal III &  0.60 &  0.20  & 0.34  \\
		 & Thermal II  		& Thermal IV &  0.12   &  0.04  & 0.36 \\
     		& Thermal III  		& Thermal IV &  0.51 &  0.26 & 0.50 \\
\hline \hline
     		    &  $\Delta R_{1.4}=-120$~m      & $\Delta R_{1.4}=-54$~m 		 &  0.40 &  0.14 & 0.33 \\
		    & $\Delta R_{1.4}=-120$~m       &   Baseline [ $R_{1.4}=11.1$~km ]  &  0.23   &  0.19  & 0.81\\
Thermal I ~~ & $\Delta R_{1.4}=-120$ ~m      &  $\Delta R_{1.4}=+116$~m 	 &  0.21   &  0.16  & 0.74\\
     		    &  $\Delta R_{1.4}=-54$~m      &    Baseline [ $R_{1.4}=11.1$~km ]   &  0.30 &  0.07  & 0.24 \\
	  	    & $\Delta R_{1.4}=-54$ ~m      &  $\Delta R_{1.4}=+116$~m  	 &  0.30  &  0.09  & 0.31 \\
		    &   Baseline [ $R_{1.4}=11.1$~km ]   &  $\Delta R_{1.4}=+116$~m  &  0.09 &  0.07  & 0.82 \\
\hline \hline
\end{tabular}
\caption{Summary of mismatches calculated using either the raw spectra (in column 4) or the 
best-fit templates (in column 5), for various model comparisons. The final column reports
the ratio of $\mathcal{M}$ calculated with the templates compared to the raw spectra.
The first section reports mismatches for models that share the
same cold \Rsmall EoS, while the second section reports mismatches for models with the
same cold \Rbig EoS. The third section reports the mismatches for models with the same
thermal treatment (Case I), but that vary in their cold EoS. All mismatches use overlaps
that have been maximized with respect 
to time and phase. Mismatches calculated using the smoothed spectral templates are typically
smaller than those calculated from the raw spectra.}
  \label{table:mismatches}
\end{table*}

\begin{figure*}[!ht]
\centering
\includegraphics[width=0.8\textwidth]{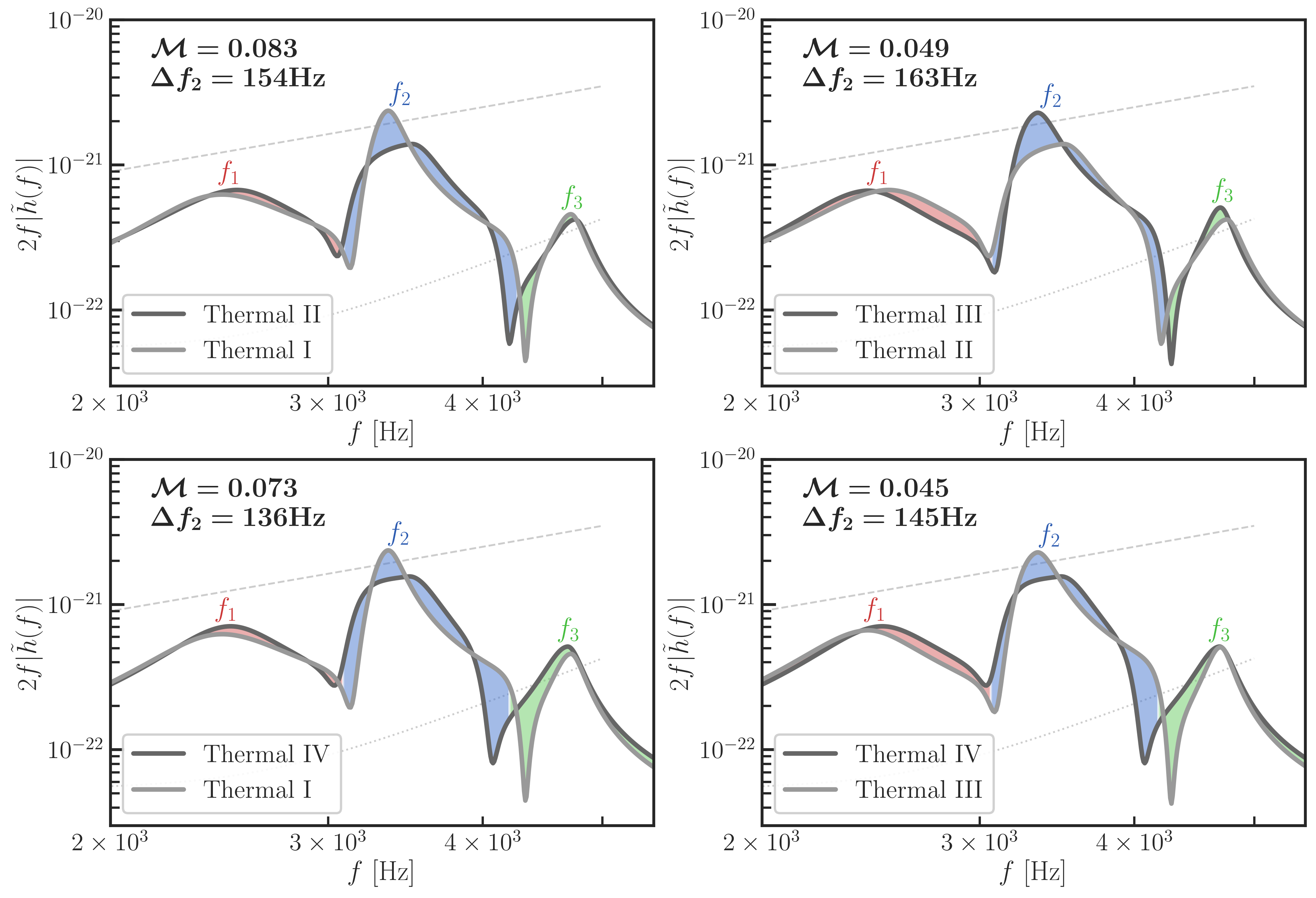}
\caption{\label{fig:R11spectra} Post-merger spectral templates for the \Rsmall EoSs that vary in their thermal treatment.
The color-coding is included to help visually distinguish the three spectral peaks (labeled $f_1$, $f_2$, and $f_3$) and
to highlight the frequency-dependence of the mismatches between each pair of spectra. The global mismatch (calculated
over the entire frequency range shown) and the difference in peak frequency are included in the top corner, for reference. All other details are identical to Fig.~\ref{fig:spectra}.}
\end{figure*}

\begin{figure*}[!ht]
\centering
\includegraphics[width=0.8\textwidth]{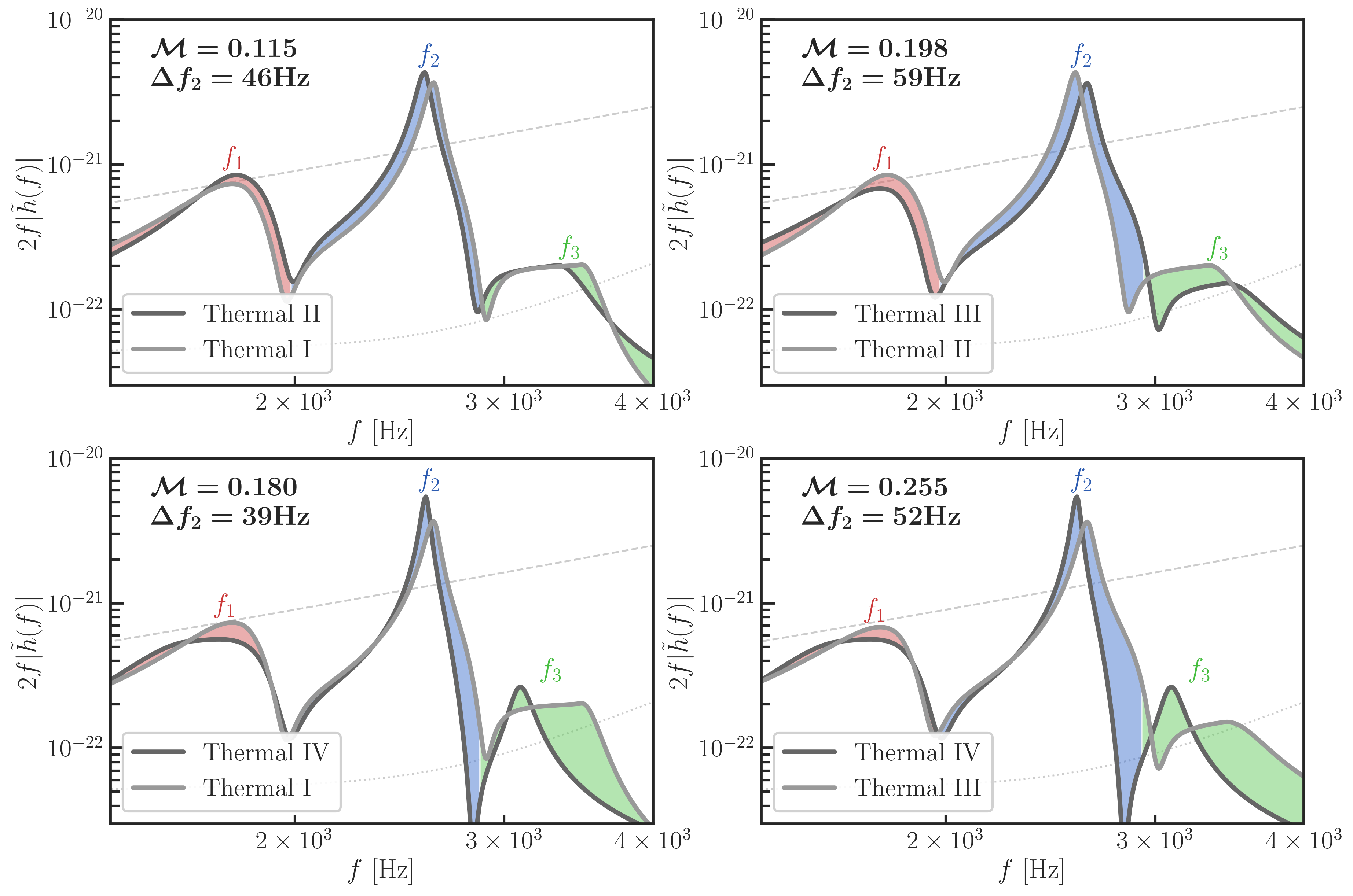}
\caption{\label{fig:R14spectra} Same as Fig.~\ref{fig:R11spectra}, but for the \Rbig EoSs.}
\end{figure*}

 Finally, for completeness, Figs.~\ref{fig:R11spectra} and \ref{fig:R14spectra} show the best-fit spectral
templates for additional examples of the \Rsmall and \Rbig EoSs with varying thermal prescriptions. These comparisons correspond to thermal models that differ in their $\alpha$ power-
law parameter (see Table~1), which determines the rate at which the effective mass function 
decays with density, and which was found in Paper I to govern the 
location of the peak frequency of the post-merger spectra.

\bibliography{inspire,non_inspire,inspire_supplement}

\end{document}